\documentclass[twocolumn,aps,showpacs,preprintnumbers,amsmath,amssymb, superscriptaddress]{revtex4-1}
\pdfoutput=1

\usepackage{graphicx}
\usepackage{dcolumn}
\usepackage{bm}
\usepackage{color}
\usepackage{xspace}

\definecolor{dblue}{rgb}{0,0,0.6}
\definecolor{dred}{rgb}{0.9,0,0}
\definecolor{dgreen}{rgb}{0,0.4,0}
\definecolor{purple}{rgb}{0.8,0,0.8}
\renewcommand{\vec}{\mathbf}
\newcommand{\braket}[2]{\xspace{\langle #1\mid #2\rangle}\xspace}

\newcommand{\ket}[1]{\xspace\ensuremath{\mid #1\rangle}\xspace}
\newcommand{\matel}[3]{\xspace{\langle #1\mid #2\mid #3\rangle}\xspace}

\begin{document}

\preprint{}

\title{Orbital Characters Determined from Fermi Surface Intensity Patterns using Angle-Resolved Photoemission Spectroscopy}
\author{X.-P. Wang}
\affiliation{Beijing National Laboratory for Condensed Matter Physics, and Institute of Physics, Chinese Academy of Sciences, Beijing 100190, China}
\author{P. Richard}
\affiliation{Beijing National Laboratory for Condensed Matter Physics, and Institute of Physics, Chinese Academy of Sciences, Beijing 100190, China}
\author{Y.-B. Huang}
\affiliation{Beijing National Laboratory for Condensed Matter Physics, and Institute of Physics, Chinese Academy of Sciences, Beijing 100190, China}
\author{H. Miao}
\affiliation{Beijing National Laboratory for Condensed Matter Physics, and Institute of Physics, Chinese Academy of Sciences, Beijing 100190, China}
\author{L. Cevey}
\affiliation{Beijing National Laboratory for Condensed Matter Physics, and Institute of Physics, Chinese Academy of Sciences, Beijing 100190, China}
\affiliation{Laboratory for Quantum Magnetism, ICMP, Ecole Polytechnique F\'{e}d\'{e}rale de Lausanne (EPFL), Switzerland}
\author{N. Xu}
\affiliation{Beijing National Laboratory for Condensed Matter Physics, and Institute of Physics, Chinese Academy of Sciences, Beijing 100190, China}
\author{Y.-J. Sun}
\affiliation{Beijing National Laboratory for Condensed Matter Physics, and Institute of Physics, Chinese Academy of Sciences, Beijing 100190, China}
\author{T. Qian}
\affiliation{Beijing National Laboratory for Condensed Matter Physics, and Institute of Physics, Chinese Academy of Sciences, Beijing 100190, China}
\author{Y.-M. Xu}
\affiliation{Materials Sciences Division, Lawrence Berkeley National Laboratory, Berkeley, California 94720, USA}
\author{M. Shi}
\affiliation{Paul Scherrer Institut, Swiss Light Source, CH-5232 Villigen PSI, Switzerland}
\author{J.-P. Hu}
\affiliation{Beijing National Laboratory for Condensed Matter Physics, and Institute of Physics, Chinese Academy of Sciences, Beijing 100190, China}
\affiliation{Department of Physics, Purdue University, West Lafayette, Indiana 47907, USA}
\author{X. Dai}
\affiliation{Beijing National Laboratory for Condensed Matter Physics, and Institute of Physics, Chinese Academy of Sciences, Beijing 100190, China}
\author{H. Ding}\email{dingh@iphy.ac.cn}
\affiliation{Beijing National Laboratory for Condensed Matter Physics, and Institute of Physics, Chinese Academy of Sciences, Beijing 100190, China}

\date{\today}

\begin{abstract}
In order to determine the orbital characters on the various Fermi surface pockets of the Fe-based superconductors Ba$_{0.6}$K$_{0.4}$Fe$_{2}$As$_{2}$ and FeSe$_{0.45}$Te$_{0.55}$, we introduce a method to calculate photoemission matrix elements. We compare our simulations to experimental data obtained with various experimental configurations of beam orientation and light polarization. We show that the photoemission intensity patterns revealed from angle-resolved photoemission spectroscopy measurements of Fermi surface mappings and energy-momentum plots along high-symmetry lines exhibit asymmetries carrying precious information on the nature of the states probed, information that is destroyed after the data symmetrization process often performed in the analysis of angle-resolved photoemission spectroscopy data. Our simulations are consistent with Fermi surfaces originating mainly from the $d_{xy}$, $d_{xz}$ and $d_{yz}$ orbitals in these materials.
\end{abstract}



\maketitle

\section{Introduction}

The spectral intensity measured by various experimental tools is modulated by matrix elements sensitive to the nature of the states probed, as well as to the experimental setup. For particular configurations, symmetry imposes some matrix elements to vanish or to reach maxima. Taking advantage of such selection rules, one can extract precious information on the probed states. For example, numerous textbooks describe how to use Raman and infrared selection rules to reveal the symmetry of phonons and other excitations. As with other probes, symmetry plays an important role in the photoemission process. It has been used often in the past to identify the nature of the electronic states of various systems \cite{Allen_PRL1990,Ronning_PRB2005, Richard_PRB2006,Santander_Nature2010,QianPRB2011,Ding_J_Phys_Condens_Matter2011,Hwang_PRB2011}.

Unfortunately, simple photoemission selection rules are restricted to a few configurations, which are not necessarily accessible with every experimental setups. Moreover, a slight sample misalignment may cause a misinterpretation of the data. In fact, the intensity variations in the momentum space often look strange and asymmetric, and they are usually neglected by ARPES experimentalists, which refer to them as the nebulous \emph{matrix element effects}. In some cases, Fermi surface mappings are symmetrized to make them look more ``natural". Despite several attempts reported previously to reproduce experimental data \cite{Bansil_PRB2005,Hwang_PRB2011,Roca_2003,Mulazzi_PRB2006}, the determination of the orbital characters in ARPES experiments is still not performed routinely, mainly due to the complexity of the calculations. A simpler and more practical approach is needed to extract useful information that is otherwise commonly sacrificed.   

In this paper, we develop a systematic but simple approach to the calculation of photoemission matrix elements in Fermi surface mappings. We apply this technique to optimally-doped Ba$_{0.6}$K$_{0.4}$Fe$_2$As$_2$, a multi-band Fe-based superconductor for which plenty of data is available in literature \cite{Richard_PoPP2011}, and to FeTe$_{0.55}$S$_{0.45}$, an Fe-chalcogenide superconductor. A precise knowledge of the determination of the orbital characters of the low-energy bands is particularly crucial in these materials, for which superconducting pairing mechanisms involving orbital fluctuations have been proposed \cite{KontaniPRL2010}. Our calculations show remarkable agreement with experimental data in multiple experimental configurations of polarization and beam orientation.

\section{Experiment}

In order to test our numerical approach, we performed ARPES experiments on high-quality single-crystals of Ba$_{0.6}$K$_{0.4}$Fe$_2$As$_2$ and FeTe$_{0.55}$Se$_{0.45}$ under various conditions. For each experimental setup, samples have been cleaved \emph{in situ} and maintained in ultra-high vacuum conditions. ARPES Fermi surface mappings were performed at the Institute of Physics, CAS, in a weakly polarized $\pi$-configuration using a MBS T1 microwave-driven helium source ($h\nu=21.2$ eV) and a VG-Scienta R4000 electron analyzer. Synchrotron-based experiments were also performed at Swiss Light Source beamline SIS and at beamline UE112\_PGM-2b of BESSY using a VG-Scienta R4000 electron analyzer mounted in $p$ and $s$ configurations, respectively. For these experiments, photons in the 20-138 eV range with different circular and linear polarizations were used. All measurements have been performed below 20 K. 

\section{Definitions and conventional selection rules}

Photoemission is a complex quantum problem which is far from easy to handle. For a simpler description, it is very convenient to decompose this process into the three steps of the so-called \emph{3-step model} \cite{Hufner_Photoemission}: (i) excitation of an electron of initial state \ket{i} into a bulk final state; (ii) travel of the excited electron towards the surface; (iii) transmission of the excited electron through the surface into a final state \ket{f} approximated by a plane wave. During the whole process, the relaxation of the remaining electrons and their interactions with the photoelectron are neglected. Within the 3-step model, the matrix element characterizing the photoemission process is given by:

\begin{equation}
 M_{if}= \matel{f}{\vec{A}\cdot\vec{r}}{i}
\end{equation}

\noindent where $\vec{A}$ is the potential vector associated with the incoming photon and $\vec{r}$ is the position operator. For sake of clarity, we also disregarded a constant prefactor. 

\begin{figure}[htbp]
\begin{center}
\includegraphics[width=8cm]{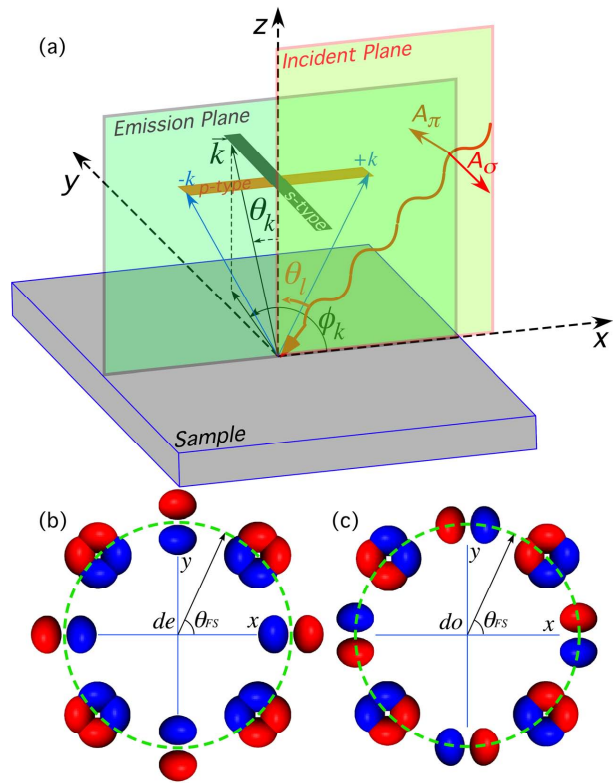}
\end{center}
\caption{\label{setup}
(Color online) (a) Definitions of the $p$ and $s$ ARPES configurations, along with the various angles used in the calculations. (b) and (c) illustrate, respectively, the even and odd combination of the $d_{xz}$ and $d_{yz}$ orbitals, along with the definition of the $\theta_{FS}$ angle.
}
\end{figure}

We present in Figure \ref{setup}(a) two commonly used ARPES configurations that simplify the analysis significantly. We call $\vec{A_{\pi}}$ and $\vec{A_{\sigma}}$, respectively, the components of the potential vector parallel ($\pi$ polarization) and perpendicular ($\sigma$ polarization) to the \emph{emission plane} defined by the vector $\vec{k}$ along which the photoemitted electron is ejected and the normal to the sample surface. Similarly, the \emph{incident plane} is defined by the incident light vector and the normal to the sample surface. When dealing with unpolarized light, it is also useful to define two special configurations of ARPES setup. Hereafter, we call $p$ and $s$ the ARPES configurations for which the incident and emission planes are parallel and perpendicular, respectively. With $\theta_l$ described as in Figure \ref{setup}(a), the potential vector can be expressed in a more general configuration with linear polarized light as:

\begin{equation}
\vec{A}= (A_x,A_y,A_z) = (-A_{\pi}\cos\theta_{l},A_{\sigma},A_{\pi}\sin\theta_{l})
\end{equation}

\noindent Right-handed circular polarization $C_+$ and left-handed circular polarization $C_-$ are defined by $\vec{A}(C_{\pm})=\vec{A_{\pi}}\pm i\vec{A_{\sigma}}$, and thus for circular polarized light we have:

\begin{equation}
\label{A_circular}
\vec{A}(C_{\pm}) = (-A_{\pi}\cos\theta_{l},\pm iA_{\sigma},A_{\pi}\sin\theta_{l})
\end{equation}

\noindent Non-polarized light is treated by adding separately the contributions of $\pi$ and $\sigma$ linearly polarized photons to the photoemission intensity $\mid M_{if}\mid^2$.

Since $\mid M_{if}\mid^2$ is a scalar observable, it must necessarily transform under crystal symmetry operations like the fully symmetric irreducible representation $\Gamma_1$ of the corresponding group in order to be different from zero. In other words, the decomposition of the tensor product of $\Gamma_i$, $\Gamma_f$ and $\Gamma_{op}$, which are the representations associated to \ket{i}, \ket{f} and $\vec{A}\cdot\vec{r}$, respectively, must contain $\Gamma_1$, which is possible only if their total parity is even. The plane wave $\braket{r}{f}=e^{i\vec{k}\cdot\vec{r}}$ is always an even state with respect to the emission plane. With respect to that same plane, the operator $\vec{A}\cdot\vec{r}$ has an even and a odd parity for light polarization parallel ($\vec{A}_{\pi}$) and perpendicular ($\vec{A}_{\sigma}$) to the emission plane, respectively. Knowing the parity of both $\vec{A}\cdot\vec{r}$ and the final state from the experimental configuration, one can deduce the parity of the initial state by choosing a proper set of coordinates. For a tetragonal system with $d$ electrons like the Fe-based superconductors, the most natural orientations for ARPES experiments is to align the sample (i) with the Fe-Fe bonds parallel to the emission plane to probe the electronic states along the $\Gamma$-M direction (here defined in the 1-Fe/unit cell representation), or (ii) with the Fe-Fe bonds at 45$^{\circ}$ from the emission plane for ARPES measurements along the $\Gamma$-X direction. In these simple cases, the five orbital wave functions $d_{z^2}$, $d_{xz}$, $d_{yz}$, $d_{xy}$ and $d_{x^2-y^2}$ form a convenient basis to describe the initial state. It is often preferable though to use linear combinations of $d_{xz}$ and $d_{yz}$ to construct the wave functions $d_o$ and $d_e$, which are odd and even with respect to any emission plane, respectively, as shown in Figures \ref{setup}(b) and (c). More specifically, we have

\begin{eqnarray}
d_e &= d_{xz}\cos\theta_{FS}+d_{yz}\sin\theta_{FS} \\
d_o &= -d_{xz}\sin\theta_{FS}+d_{yz}\cos\theta_{FS} 
\end{eqnarray}

\noindent where $\theta_{FS}$ is the Fermi surface angle defined in Figures \ref{setup}(b) and (c). Although such approach has been used already to study the Fe-based superconductors \cite{FinkPRB2009,MalaebJPSJ2009,Y_XiaPRL2009,MansartPRB2011,Nishi_PRB2011,Y_Zhang_PRB2011}, the various interpretations are not always consistent, thus calling for alternative methods for determining the orbital characters.

\section{Computational details}\label{computational_details}

In this section, we explain briefly how to use ARPES intensity patterns to determine the orbital characters of the Fe $3d$ electronic states near the Fermi level of Fe-based superconductors. A more detailed calculation is given in Appendix A. Here we focus only on the main steps. 

Within the 3-step model, as mentioned previously, we use the $3d$ atomic orbital wave functions $\{d_{xy},d_{xz},d_{yz},d_{z},d_{x^2-y^2}\}$ to characterize the initial state \ket{i}:

\begin{equation}
\braket{\vec{r}}{i} = R_{32}(r)\sum_{m=-2}^{2}\alpha_mY_2^{m}(\theta,\phi)
\end{equation}

\noindent where $\alpha_m$ are coefficients, $R_{32}(r) \propto r^2e^{-r/3}$ is the longitudinal part of the $3d$ atomic wave functions with $r$ given in Bohr radius units and $Y_l^{m}(\theta, \phi)$ is the spherical harmonic with angular moment $l$ and azimuthal moment $m$. The final state, here approximated by a plane wave function, can be expressed in terms of the spherical harmonics as:

\begin{eqnarray}
	\braket{\vec{r}}{f}&=&e^{i\vec{k_f}\cdot \vec{r}}\nonumber\\
	=4\pi\sum_{l=0}^{\infty}&i^l&j_l(k_fr)\sum_{m=-l}^lY_l^{m*}(\theta_k, \phi_k) Y_l^m(\theta,\phi)
\end{eqnarray}

\noindent where $j_l(k_fr)$ is the Bessel function. The photoemission matrix element $M^{\lambda}_{if}$ associated with the different spherical harmonic $Y_2^{\lambda}$ becomes

\begin{eqnarray}
M_{if}^{\lambda} &\propto& \matel{f}{\vec{A}\cdot\vec{r}}{i;m=\lambda}\nonumber
\\
&=&(A_x\Upsilon_x^\lambda + A_y\Upsilon_y^\lambda  + A_z\Upsilon_z^\lambda )
\end{eqnarray}

\noindent where, 

\noindent\begin{eqnarray}
\Upsilon_{\alpha=x,y,z}^{m=\lambda}\notag\\
=\sum_{l=0}^{\infty}i^l&\rho_l(k_f)\displaystyle\sum_{\mu=-l}^lY_l^{\mu}(\theta_k,\phi_k)f_{\alpha}^{\lambda}(l,\mu)
 \label{eq:Matorbital} 
\end{eqnarray}

\begin{eqnarray}
\rho_l(k_f) &=&4\pi \int dr r^3 R_{32}(r)j_l(k_fr)
 \label{eq:MatRadial} 
\end{eqnarray}

\begin{eqnarray}
f_{\alpha}^{\lambda}(l,\mu) &=&\oint d\theta d\phi\sin\theta Y_l^{\mu*}(\theta,\phi)p_\alpha Y_2^{\lambda}(\theta,\phi)
 \label{eq:MatG3} 
\end{eqnarray}

\begin{eqnarray}
p_x&=&x/r=\sqrt{\frac{1}{2}}(Y_1^{-1}-Y_1^1)\notag\\
p_y&=&y/r=i\sqrt{\frac{1}{2}}(Y_1^{-1}+Y_1^1)\notag\\
p_z&=&z/r=Y_1^0
\end{eqnarray}

The passage from these matrix elements to matrix elements involving the $3d$ orbital atomic wave functions is performed using the following relations:

\begin{eqnarray}
 M_{if}^{d_{z^2}}  &=& M_{if}^{0}\nonumber
 \\
 M_{if}^{d_{yz}}  &=& i\sqrt{\frac{1}{2}}(M_{if}^{-1}+M_{if}^{1})\nonumber
\\
 M_{if}^{d_{xz}}  &=& \sqrt{\frac{1}{2}}(M_{if}^{-1}-M_{if}^{1})\nonumber
\\
 M_{if}^{d_{xy}}  &=&  i\sqrt{\frac{1}{2}}(M_{if}^{-2}-M_{if}^{2})\nonumber
\\
 M_{if}^{d_{x^2-y^2}}  &=&  \sqrt{\frac{1}{2}}(M_{if}^{-2}+M_{if}^{2})
  \label{eq:MatO1} 
 \end{eqnarray}

\begin{figure}[htbp]
\begin{center}
\includegraphics[width=8cm]{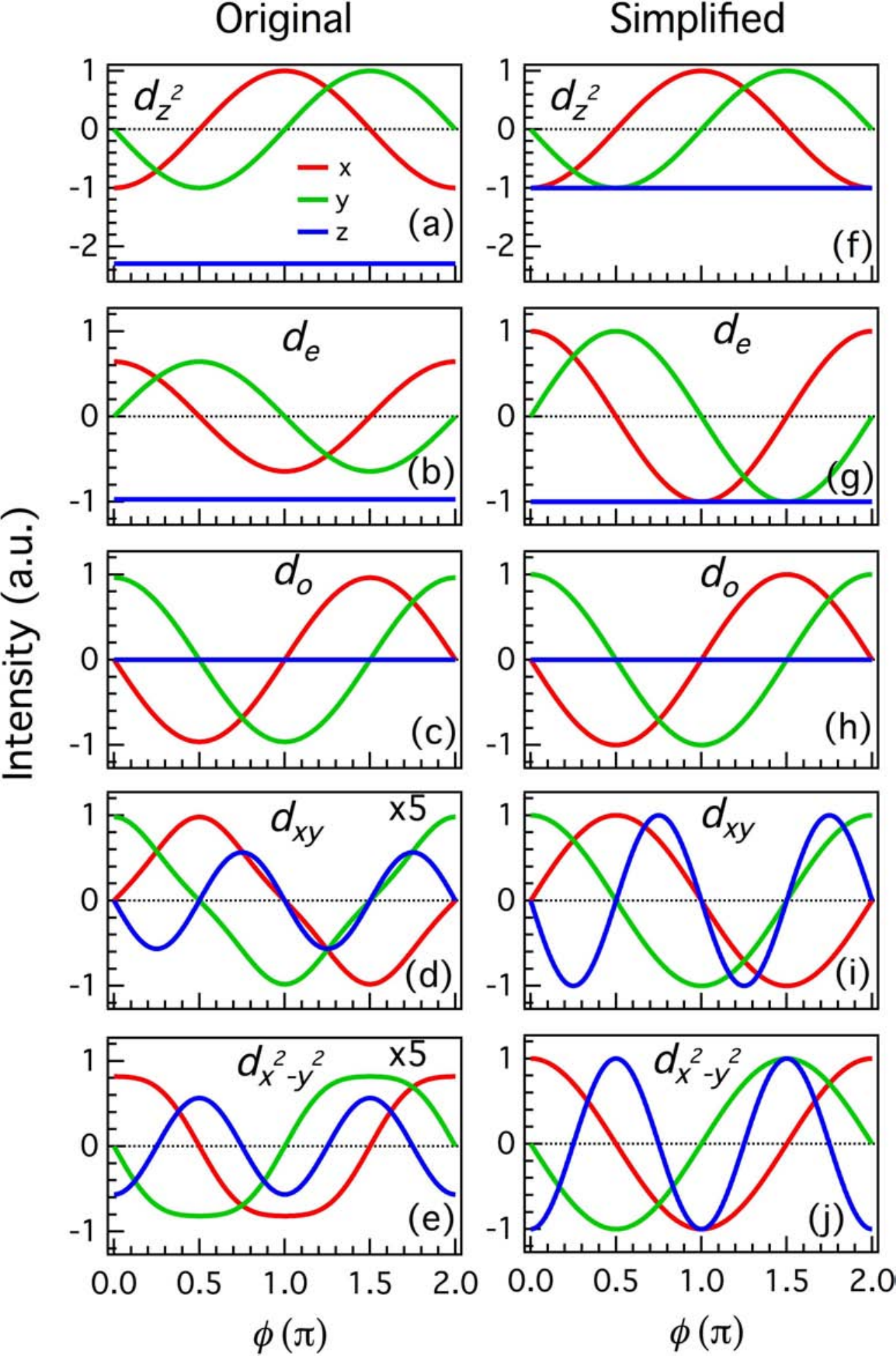}
\end{center}
\caption{
\label{xyzMatrix}
(Color online) (a)-(e) Angular dependence of the $x$ (in red), $y$ (in green) and $z$ (in blue) components of the photoemission matrix elements related to the various 3$d$ orbitals (see the text). (f)-(j) Same as (a)-(e) but using the simplified matrix elements (see the text). We used $h\nu=21.2$ eV and $k_F = 0.3\pi$.}
\end{figure}

In Figures \ref{xyzMatrix}(a)-(e), we give the $\phi$ dependence of the $x$, $y$ and $z$ components of these matrix elements for a photon energy of 21.2 eV, which corresponds to the I$\alpha$ line of conventional He discharge lamps, and for $k_{f||}=0.3\pi/a$, where $a$ is the in-plane lattice parameter. We used the fact that the standard Gaunt coefficients $f_{\alpha}^{\lambda}(l,\mu)$ are non-vanishing only for $l=1,3$. In addition, we found empirically that the coefficients $\rho_3(k_f)\ = 1$ and $\rho_1(k_f)\ = -2/5$ reproduce the experimental data very well over a wide range of photon energy. For a better comparison, all the matrix element weights are normalized by $z-M(d_e)$. We note that matrix elements for the purely in-plane orbitals $d_{xy}$ and $d_{x^2-y^2}$ are smaller than the other ones by a factor of 5, even though $d_{xy}$, $d_{xz}$ and $d_{yz}$ are equivalent orbitals under symmetry operations. This effect is caused by the smallness of the angle $\theta_k$ when using a photo energy in this range. We also point out that the $z$ component of the $d_{z^2}$ matrix element is larger than any other, which indicates that the $d_{z^2}$ matrix element is more sensitive than others to a $A_z$ polarization. 

Due to the fact that $k_z$ is not a good quantum number in photoemission experiments, we introduce a few empirical parameters to the formula describing the full matrix element M$^{\delta}_{if}$ and improve the agreement between simulations and experimental data. The full matrix element is now expressed as:

\begin{eqnarray}
\vec{M}_{if}^{\delta}=(A_xM_{x}^{\delta}+A_yM_{y}^{\delta}+A_zM_{z}^{\delta}*w_z*e^{i\gamma_\delta})w_\delta
\label{eqMif}
\end{eqnarray}

\noindent where $\delta = d_{z^2}, d_{xz},d_{yz},d_{xy},d_{x^2-y^2}$. From direct comparison with experiments, we found out that the ratio between the matrix elements associated with the $d_{xy}$ and $d_{yz}$ orbitals is only 1/2 instead of 1/5. We thus introduced the weight factor $w_{\delta}=\{5 (\delta=d_{xy},d_{x^2-y^2}), 1$ otherwise\}. Within our semi-quantitative approach, these parameters are viewed as phenomenological parameters compensating for our simplified model. They have been fixed at the same values for all our simulations. For the study of Fe-based superconductors, $w_z$ and $\gamma$ have been fixed to 4 and $k_zc+\frac{3\pi}{2}$, respectively, where $c$ is the lattice parameter along the $z$ direction. 


Since ARPES allows only measurement of the intensity $I_{if}^{\delta}(\phi_k)=|M^{\delta}_{if}|^2$, and more precisely the relative distribution of intensity in the momentum space, several prefactors can be dropped in the calculations, including the imaginary prefactor. Assuming small $\theta_k$ in Eq.~(\ref{eq:Matorbital}), we can simplified the matrix elements as follows, keeping only the angular parts of the matrix element components:

\begin{eqnarray}
&M(d_{z^2})&=(-\cos\phi_{k}, -\sin\phi_{k},-1)\nonumber
\\
&M(d_{xz})&=(1, 0,-\cos\phi_{k})\nonumber
\\
&M(d_{yz})&=(0, 1,-\sin\phi_{k})\nonumber
\\
&M(d_{xy})&=(\sin\phi_{k}, \cos\phi_{k},-\sin2\phi_{k})\nonumber
\\
&M(d_{x^2-y^2})&=(\cos\phi_{k}, -\sin\phi_{k},-\cos2\phi_{k})\nonumber
\\
&M(d_{e})&=M(d_{xz})\cos\theta_{FS} + M(d_{yz})\sin\theta_{FS} \nonumber
\\
&M(d_{o})&=-M(d_{xz})\sin\theta_{FS} + M(d_{yz})\cos\theta_{FS} \nonumber
\end{eqnarray}

The different components of these simplified matrix elements are given in Figures \ref{xyzMatrix}(f)-(j). Although their precise absolute values differ from that of the components in Figures \ref{xyzMatrix}(a)-(e), they carry essentially the same orbital information while simplifying calculations significantly.

We now consider the effect of light polarization on the photoemission response, which is widely known by experimentalists to be important. We first start by the experimental observation of a difference, often called circular dichroism, between the photoemission response to left-handed and right-handed circular polarizations. This effect can have different origins \cite{Venus_PRB1993}. For example, it has been attributed to spontaneous breaking of the time-reversal symmetry in Bi$_2$Sr$_2$CaCu$_2$O$_{8+\delta}$ \cite{Kaminski_Nature2002}. This effect is quite different from the circular polarization used in YBa$_2$Cu$_3$O$_{7-\delta}$, for which circular dichroism appears as a surface anomaly. It has been useful to separate the photoemission contributions of the bulk and of the highly polar surface resulting from the absence of natural cleaving plane in this material \cite{Nakayama_PRB2007, Zabolotnyy_PRB2007_2, Nakayama_PRB2009}. In this particular case, only a non-trivial combination of the photoemission responses to left-handed and right-handed circular polarized light can allow a full separation of these two components \cite{Zabolotnyy_PRB2007_1}. 

We note that besides these \emph{anoumalous} circular dichroism effects, one should also expect asymmetric photoemission response to $C_{+}$ and $C_{-}$ light, depending on the geometry of the ARPES configuration. Indeed, it has been shown that unless the photoemitted momentum, the normal to the sample surface and the incident beam momentum are all coplanar ($p$ ARPES configuration) in a mirror symmetry plane of the sample, circular dichroism can be observed \cite{Venus_PRB1993}. Rather than searching an origin for circular dichroism in the Fe-based superconductors, which goes beyond the purpose of the current work, \emph{i.e.} to extract useful information on the orbital characters of the bands and FSs observed by ARPES, here we simply try to describe its phenomenology and to add it as a tool to determine the orbital characters of bands.  
 
By working out the details of Eq.~(\ref{eq:MatO1}), one can show that all the matrix elements $M_{\alpha=x,y,z}^{{\delta}=d_{z^2}, d_{xz}, d_{yz}, d_{xy}, d_{x^2-y^2}}$ have pure imaginary values. Assuming the form of the potential vector given in Eq. \eqref{A_circular} for circular polarization, we deduce that the photoemission intensity for $C_{\pm}$ polarized light is given by: 

\begin{eqnarray}
\label{eq:Intcpcm} 
I_{C_{\pm}}^\delta&\propto& \vert M_{C_{\pm}}^\delta\vert ^2\notag\\
&\propto& \vert A_{\pi}(-\cos\theta_{l}M_x^\delta+\sin\theta_{l}M_z^\delta) \pm iA_{\sigma}M_y^\delta\vert^2\\
&=&A_{\pi} ^2\vert -\cos\theta_{l}M_x^\delta+\sin\theta_{l}M_z^\delta\vert^2 + A_{\sigma}^2\vert M_y^\delta\vert^2\notag
\end{eqnarray}

The previous equation indicates that there should be no difference between the photoemission responses to $C_+$ and $C_-$ polarized light if all the matrix elements have pure imaginary values, which is supposed from Eq. (\ref{eq:MatO1}). To account for the difference occurring in real experimental data, we add a phase to each matrix element. The photoemission intensity when circular polarized light is used thus becomes:

\begin{eqnarray}
I_{C_{\pm}}^\delta\propto \vert A_{\pi}(-\cos\theta_{l}M_x^\delta e^{i\gamma_x}+\sin\theta_{l}M_z^\delta e^{i\gamma_z})\pm iA_{\sigma}M_y^\delta e^{i\gamma_y}\vert^2\notag\\
 \label{eq:Matmod} 
\end{eqnarray}

\noindent The addition of phase factors to each matrix element also influences the photoemission intensity $I_{\pi}$ corresponding to $\pi$-polarized light and the photoemission intensity $I_{\sigma}$ associated to $\sigma$-polarized light, which are now respectively expressed as: 
\begin{eqnarray}
I_{\pi}^\delta\propto \vert A_{\pi}(-\cos\theta_{l}M_x^\delta e^{i\gamma_x}+\sin\theta_{l}M_z^\delta e^{i\gamma_z})\vert^2
 \label{eq:Intp} 
\end{eqnarray}

\noindent and

\begin{eqnarray}
I_{\sigma}^\delta\propto \vert A_{\sigma}M_y^\delta e^{i\gamma_y}\vert^2
 \label{eq:Intp0} 
\end{eqnarray}

These later considerations allow us to predict appropriate phenomenological forms for the photoemission intensity responses to circular and unpolarized light in the common $p$ and $s$ ARPES configurations. As illustrated in Figure \ref{setup}(a), the photoemitted electrons in the $p$-type ARPES configuration are collected in the $k_x-k_z$ plane. For odd symmetry orbital characters, we then have $M_x^{\delta=odd}=0$, $M_z^{\delta=odd}=0$, but $M_y^{\delta=odd}\neq0$, while for even symmetry we get $M_x^{\delta=even}\neq0$, $M_z^{\delta=even}\neq0$, but $M_y^{\delta=even}=0$. As a consequence, 

\begin{eqnarray}
I_{C_{\pm}}^{\delta=odd}({\pi})&\propto& \vert\pm iA_{\sigma}M_y^\delta e^{i\gamma_y}\vert^2\notag\\
I_{C_{\pm}}^{\delta=even}({\pi})&\propto& \vert A_{\pi}(-\cos\theta_{l}M_x^\delta e^{i\gamma_x}+\sin\theta_{l}M_z^\delta e^{i\gamma_z})\vert^2\notag
\label{eq:Intpodd} 
\end{eqnarray}

From these later equations, we can conclude that in the $p$ configuration, there is no difference between $C_+$ and $C_-$ along the high symmetry line cut, a result valid for both odd and even orbital characters and consistent with a previous work \cite{Venus_PRB1993}. This contrasts with the intensity predicted for a non-polarized light excitation. Since the photoemission intensity $I_{non}$ for a non-polarized light excitation can be described by the sum of  $I_{\pi}$ and $I_{\sigma}$, we have:

\begin{eqnarray}
I_{non}^{\delta=odd}({\pi})&\propto& \vert A_{\sigma}M_y^\delta e^{i\gamma_y}\vert^2\notag\\
I_{non}^{\delta=even}({\pi})&\propto& \vert A_{\pi}(-\cos\theta_{l}M_x^\delta e^{i\gamma_x}+\sin\theta_{l}M_z^\delta e^{i\gamma_z})\vert^2\notag
 \label{eq:Intpnon} 
\end{eqnarray}

These equations indicate that in the $p$ ARPES configuration, even symmetry orbitals may lead to an intensity asymmetry along $k_x$, but not the odd symmetry ones.

In the $s$ configuration, electrons are collected in the $k_y-k_z$ plane, and we should expect different selections rules. Indeed, we now have for the odd symmetry orbital characters: $M_x^{\delta=odd}\neq0$, but $M_z^{\delta=odd}=0$ and $M_y^{\delta=odd}=0$. For the even symmetries, the experimental configuration imposes $M_x^{\delta=even}=0$, but $M_z^{\delta=even}\neq0$ and $M_y^{\delta=even}\neq 0$. Consequently, the photoemission intensity response to circular polarized light in the $s$ configuration becomes:

\begin{eqnarray}
I_{C_{\pm}}^{\delta=odd}({\sigma})&\propto&\vert A_{\pi}(-\cos\theta_{l}M_x^\delta e^{i\gamma_x})\vert^2\notag\\
I_{C_{\pm}}^{\delta=even}({\sigma})&\propto& \vert A_{\pi}(\sin\theta_{l}M_z^\delta e^{i\gamma_z})\pm iA_{\sigma}M_y^\delta e^{i\gamma_y}\vert^2\notag
 \label{eq:Intsodd} 
\end{eqnarray}

In contrast to the $p$ configuration, the equations show that we can expect circular dichroism in the $s$ configuration, in agreement with a previous work focused on core levels \cite{Venus_PRB1993}. The use of circular polarized light is also a useful way to determine the symmetry of the band structure. As for the photoemission response to non-polarized light in the $s$ ARPES configuration, we now have:

\begin{eqnarray}
I_{non}^{\delta=odd}({\sigma})&\propto& \vert A_{\pi}(-\cos\theta_{l}M_x^\delta e^{i\gamma_x})\vert^2\notag\\
I_{non}^{\delta=even}({\sigma})&\propto& \vert A_{\pi}(\sin\theta_{l}M_z^\delta e^{i\gamma_z})\vert^2+\vert A_{\sigma}M_y^\delta e^{i\gamma_y}\vert^2\notag
 \label{eq:Intsnon} 
\end{eqnarray}

Even though the comparison between the photoemission intensity recorded with linear $\pi$-polarized and $\sigma$-polarized light give the strongest contrasts, some assumptions on the symmetry of bands can still be made based on data recorded with non-polarized light, such as a traditional He discharge lamp.

\section{Orbital characters in $\textrm{Ba}$$_{0.6}$K$_{0.4}$$\textrm{Fe}$$_2$A$s_2$}\label{section122}

Following LDA band calculations indicating that the orbital weight around the Fermi level in Ba$_{0.6}$K$_{0.4}$Fe$_2$As$_2$ is dominated by the Fe $3d$ orbitals $d_{xz}$, $d_{yz}$ and $d_{xy}$, we only considered the related matrix elements in our simulations. More specifically, LDA predicts that there are three holelike Fermi surfaces centered at the $\Gamma$ point with $d_{xy}$, $d_e$ and $d_o$ orbital characters. Previous ARPES results also show the existence of 3 holelike Fermi surface pockets centered at $\Gamma$, two of them being nearly degenerate \cite{Ding_EPL2008,L_ZhaoCPL2008,Ding_J_Phys_Condens_Matter2011,YM_Xu_NPhys2011}. Following a previous notation, here we call $\beta$ the outer Fermi surface, and $\alpha$ and $\alpha'$ the two others, which will be considered degenerate in our simulations. At M $=(\pi,0)$, here defined in the 1 Fe/unit cell description, theoretical calculations predict a Fermi surface pattern formed by the hybridization of 2 ellipses. For $k_z=0$, the ellipse tips have a $d_{xy}$ orbital character while the inner part comes from $d_{yz}$ and $d_{xz}$ \cite{Graser_NJP2009,Graser_PRB2010}. This orbital distribution around M is reversed for $k_z=\pi$. Theoretical calculations also predict a non-negligible $k_z$ variation at the M point \cite{Graser_NJP2009,Graser_PRB2010,CH_Lin_PRL2011} that is not observed by ARPES \cite{YM_Xu_NPhys2011}. While ARPES performed for several Fe-based materials with different cleaving surfaces reveal $k_z$ variations of the electronic band structure at the $\Gamma$ point \cite{Richard_PoPP2011}, the reasons behind this experiment \emph{vs} theory discrepancy for the electronic band structure at the M point are still under intense debate. In our simulations, we use as M-centered electronlike Fermi surface pockets the $k_z$-invariant hybridized functions determined from a three-band model \cite{XGWen_3orbital}: 

\begin{eqnarray}
(d_{xy/yz})=&&d_{yz}t'_3\cdot i \sin\theta_{FS}  \nonumber\\
&& -d_{xy}(t'_2\cdot \cos^2\theta_{FS}+\epsilon_{xy}^0) 
\\
(d_{xz/xy})=&&d_{xz}t'_3\cdot i\cos\theta_{FS} \nonumber\\
&&-d_{xy}({t'_2\cdot \sin^2\theta_{FS}+\epsilon_{xy}^0})
\end{eqnarray}

\noindent where we imposed $t'_3=t'_2=1$ and $\epsilon_{xy}^0 = 0.1$ for convenience, these parameters making the weight of different orbital characters similar to random phase approximation (RPA) results. As we show below, such functions are at least consistent with the ARPES observations.

\begin{figure*}[!t]
\begin{center}
\includegraphics[width=16cm]{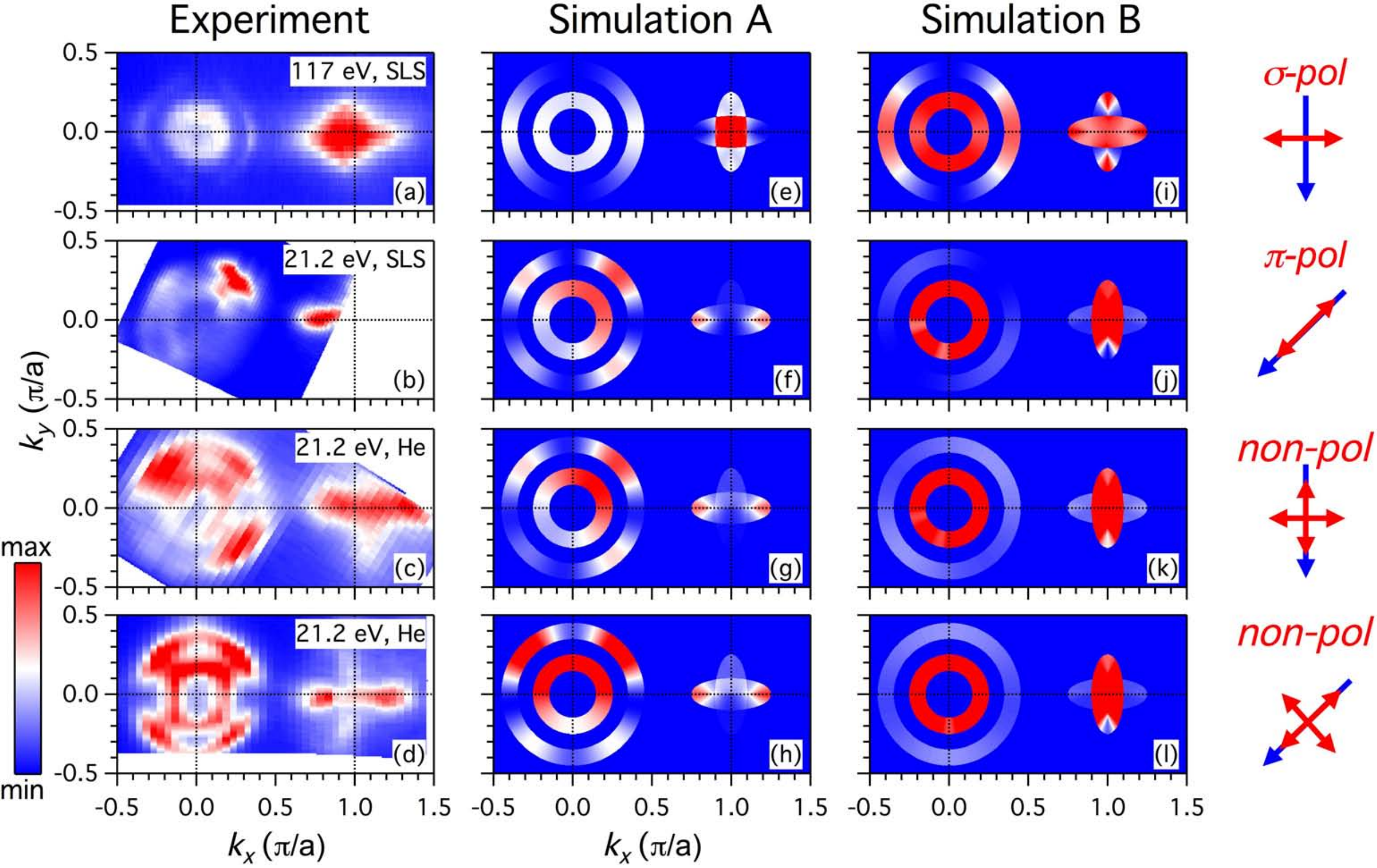}
\end{center}
\caption{\label{pattern122}
(Color online) Fermi surface intensity patterns of Ba$_{0.6}$K$_{0.4}$Fe$_2$As$_2$. (a)-(d) Experimental results with different photon energies, polarizations and incident beam directions. (e)-(h) Corresponding simulated results using the simplified model described in the text (Simulation A: optimized orbital configuration). The inner $\Gamma$-centered $\alpha$ and $\alpha'$ Fermi surface pockets with $d_e$ and $d_o$ orbital characters are considered degenerate. The outer one ($\beta$ band) is associated to the $d_{xy}$ orbital. The tip of the M-centered Fermi surface pockets has pure $d_{xz}$ or $d_{yz}$ orbital characters while the inner part carries a dominant $d_{xy}$ orbital character. (i)-(l) Same as (e)-(h) but using a wrong orbital assignment (Simulation B). The orbital characters of the $\beta$ and $\alpha'$ bands have been exchanged compared to Simulation A. The orbital characters of the tip and inner part of the M-centered FS have also been exchanged. Red double-arrows and blue arrows indicate the in-plane components of the orientation of the light polarization and direction, respectively.
}
\end{figure*}

Figures \ref{pattern122}(a)-(d) show the Fermi surface intensity patterns of Ba$_{0.6}$K$_{0.4}$Fe$_2$As$_2$ in various configurations. For each experimental pattern, we give in the second column from the left the corresponding result from our calculations using the orbital configuration given above (Figures \ref{pattern122}(e)-(h): Simulation A). The size of each Fermi surface used in the calculations is chosen to match approximately the size of the corresponding experimental Fermi surface. We note that small variations in the Fermi surface sizes do not have qualitative effect on the calculated patterns. In the first experimental configuration, light is $\sigma$-polarized along the $x$-axis direction. The experimental results indicate much stronger weight at the M point than for the $\Gamma$-centered Fermi surfaces. The intensity is even weaker for the $\beta$ band, especially along the $k_y$ direction. Since the polarization is parallel to $k_x$, this result suggests that the $\beta$ band must have an odd symmetry along both $k_x$ and $k_y$, and we thus tentatively associate the $\beta$ band to a $d_{xy}$ orbital character, leaving $d_o$ and $d_e$ symmetries for the nearly degenerate $\alpha$ and $\alpha'$ bands. In this configuration, our simulation shows a much stronger intensity at the M point than for the $\Gamma$-centered Fermi surfaces, in agreement with experimental data. Moreover, it predicts weaker intensity on the outer $\Gamma$-centered Fermi surface, with even weaker spectral intensity along the $k_y$ direction, which is also consistent with the experiment. 


To test our approach and our orbital attributions further, we show in Figure \ref{pattern122}(b) results obtained at 21.2 eV (near $k_z=0$ \cite{YM_Xu_NPhys2011}) with light $\pi$-polarized along $\Gamma$-X$ (\pi/2,\pi/2)$. The Fermi surface mapping is quite counter-intuitive, with very strong intensity spots found on the $\Gamma$-centered Fermi surface pockets in the first quadrant. The result also indicates strong intensity on the tip of the ellipse that has been measured. Surprisingly, even such a peculiar Fermi surface pattern is qualitatively well reproduced by our simulation displayed in Figure \ref{pattern122}(f), except perhaps for a weaker intensity on the inner $\Gamma$-centered bands than expected. This good agreement between simulation and experiment reinforces our initial orbital assignment.   

In Figures \ref{pattern122}(c) and (d), we present the Fermi surfaces obtained with non-polarized light from the I$\alpha$ spectral line of a Helium discharge lamp ($h\nu=21.2$ eV) for a beam incidence aligned along the $\Gamma$-X and $\Gamma$-M directions, respectively. Although both configurations give rise to much stronger intensity along the M-centered Fermi surface elongated along $k_x$ than the one elongated along $k_y$, the Fermi surface patterns are quite different around the Brillouin zone center. While the map obtained with the $\Gamma$-X orientation of the light shows spectral intensity almost suppressed in the third quadrant, the intensity has a more symmetric distribution in the map recorded in the $\Gamma$-M configuration, albeit for an intensity slightly smaller below the $k_x=0$ line than above. Moreover, the $\beta$ Fermi surface exhibits an additional suppression of intensity along $k_x$ and $k_y$. Once more, our simulations, displayed in Figures \ref{pattern122}(f) and (g), explain well the strange spectral weight intensity distribution found experimentally. 
 
\begin{figure}[!t]
\begin{center}
\includegraphics[width=9cm]{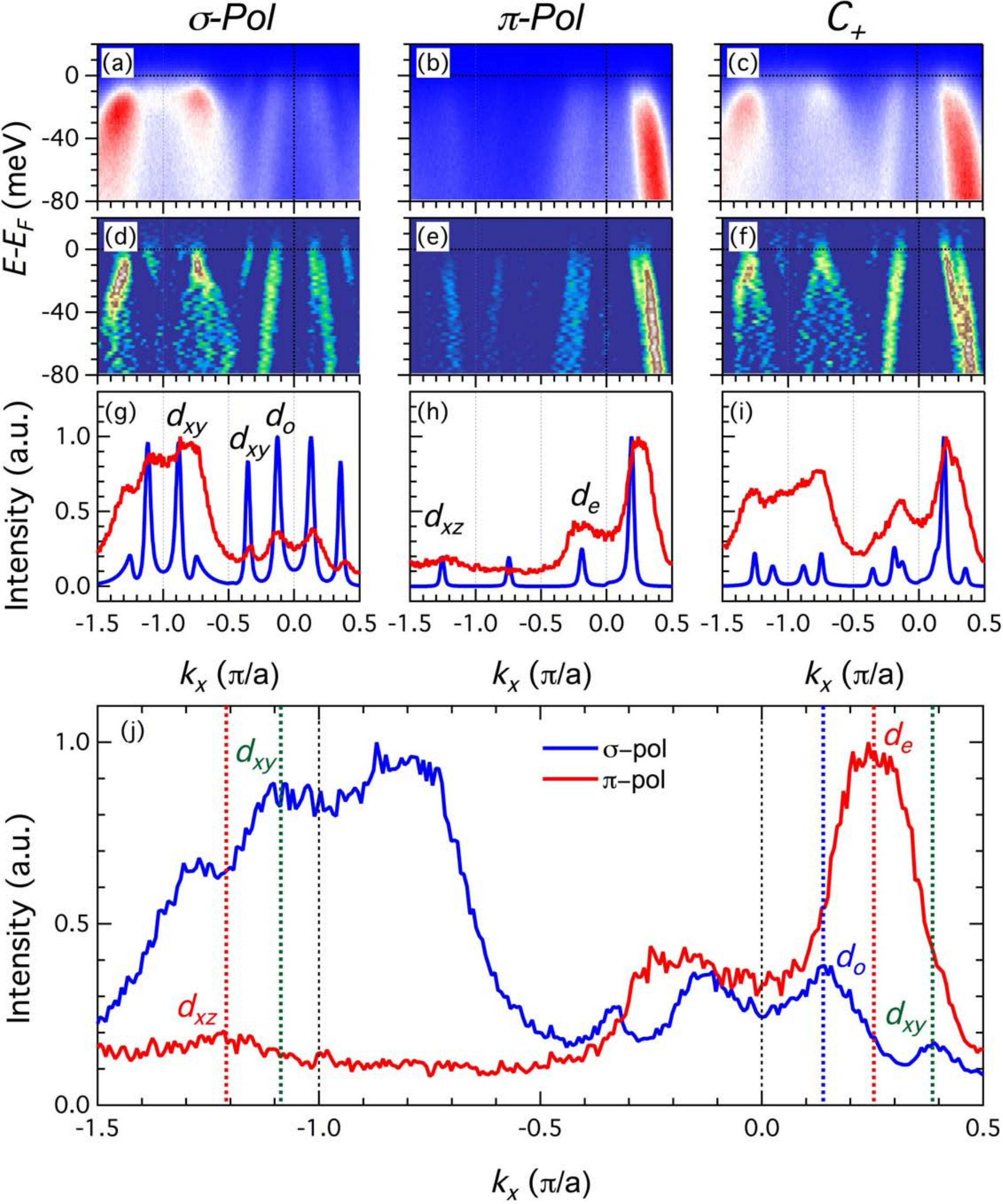}
\end{center}
\caption{\label{MDC_profiles}
(Color online) 
First row: ARPES intensity plots of Ba$_{0.6}$K$_{0.4}$Fe$_2$As$_2$ recorded along the M-$\Gamma$ direction using 138 eV photons and (a) $\sigma$, (b) $\pi$ and (c) circular right polarizations. Second row: corresponding 1D curvature intensity plots \cite{Zhang_RSI2011} along the momentum direction. Third row: corresponding MDC profile integrated within 10 meV below $E_F$, compared to profiles simulated using the same half-width at half-maximum for each band. (j) Direct comparison of the MDC profiles recorded with $\sigma$ (red) and $\pi$ (blue) polarizations.}
\end{figure}

In the third column from the left in Figure \ref{pattern122}, we illustrate the sensitivity of our approach to distinguish between two sets of simulations by displaying simulation results (Simulation B) using a \emph{wrong} orbital assignment. As compared to Simulation A, we exchanged the orbital characters of the $\alpha'$ ($d_o$ in Simulation A) and $\beta$ ($d_{xy}$ in Simulation A) bands. We also switched the orbital characters around the M point, where now the tip is considered to have a $d_{xz}/d_{yz}$ character as opposed to a $d_{xy}$ orbital character for the inner part. Although the results seem also good when using $\sigma$-polarized light, the agreement becomes much worst for other configurations. This observation is valid not only for the $\beta$ band, but also for the Fermi surface intensity pattern at the M point, which is mainly aligned along $k_y$ rather than $k_x$, in contrast to the experimental results and to Simulation A. For these reasons, we argue that the orbital configuration used in Simulation A is at least compatible with the experimental results, whereas the one used in Simulation B must be discarded. 

Additional information can be obtained from the simulations away from the Fermi level. In Figure \ref{MDC_profiles} we display the ARPES intensity plots of Ba$_{0.6}$K$_{0.4}$Fe$_2$As$_2$ recorded along the M-$\Gamma$ direction using 138 eV photons. This photon energy corresponds to $k_z=\pi$, where the $\alpha$ and $\alpha'$ bands have the largest separation and thus their apparent degeneracy is removed \cite{YM_Xu_NPhys2011}. As expected, the intensity pattern is strongly polarization-dependent. While the M-centered bands have very high intensity for $\sigma$ polarization as compared with the $\Gamma$-centered bands, the opposite is observed for $\pi$ polarization. The spectrum obtained with circular polarization is more or less an hybrid of the two others. Interestingly, the spectral weight is strongly asymmetric with respect to the zone center when using $\pi$ polarization, whereas it is almost symmetric for the spectrum recorded with $\sigma$ polarization. 

The dispersions and Fermi wave vectors of the various bands can be approximated from the intensity plots as well as from the corresponding curvature intensity plots \cite{Zhang_RSI2011}, which are given in the second row of Figure \ref{MDC_profiles}. Using this information, we performed simulations for energies away from the Fermi level. The results are compared directly to the MDC profiles in Figure \ref{MDC_profiles}. To simplify, we attributed the same half-width at half-maximum to each band. Despite this simplification, the simulations allow a good understanding of the MDC profiles. For example, the simulations predict the relative symmetry and asymmetry of the photoemission intensity with respect to the zone center. More importantly, they allow us to pin down the orbital characters of the $\alpha$ and $\alpha'$ bands. Symmetry imposes the intensity of the $d_e$ band to vanish when using $\sigma$ polarization along that particular direction. Accordingly, only two bands are observed around $\Gamma$ in this configuration. In contrast, both the $d_{xy}$ and $d_o$ bands should vanish around $\Gamma$ when using $\pi$ polarization. Accordingly, only one band is detected around $\Gamma$ using $\pi$ polarization. Since these bands have different Fermi wave vectors, their orbital characters appear clearly after superimposition of the MDC profiles of the spectra recorded with the $\sigma$ and $\pi$ polarizations, as illustrated in Figure \ref{MDC_profiles}(j). For instance, we conclude that while the innermost band, the $\alpha$ band, has a $d_o$ symmetry, the $\alpha'$ band corresponds to the even combination of the $d_{xz}$ and $d_{yz}$ orbitals. Our simulations also confirm that the $\beta$ band carries a dominant $d_{xy}$ orbital character. 

Even though the situation is a little more complicated around the M point due the weaker photoemission intensity with $\pi$-polarized light, our simulations reproduce qualitatively well the experimental MDC profiles given in Figures \ref{MDC_profiles}(g), \ref{MDC_profiles}(h) and \ref{MDC_profiles}(i), and suggest that the orbital character at the tip of the electronlike Fermi surface pockets with ellipsoidal shape is $d_{xz}$ ($d_{yz}$). This conclusion differs from a previous ARPES study on Co-doped BaFe$_2$As$_2$ that rather attributed $d_{x^2-y^2}$ and $d_{z^2}$ characters to the tip \cite{Y_Zhang_PRB2011}, which does not show up at the M point in LDA band calculations \cite{Zhang_CPL2009,Vildosola_2008,CH_Lin_PRL2011}. However, both ARPES studies indicate that the shape and orbital characters of the Fermi surfaces at the M point is preserved along $k_z$, in contrast to LDA band calculations. 


\begin{figure}[!t]
\begin{center}
\includegraphics[width=8cm]{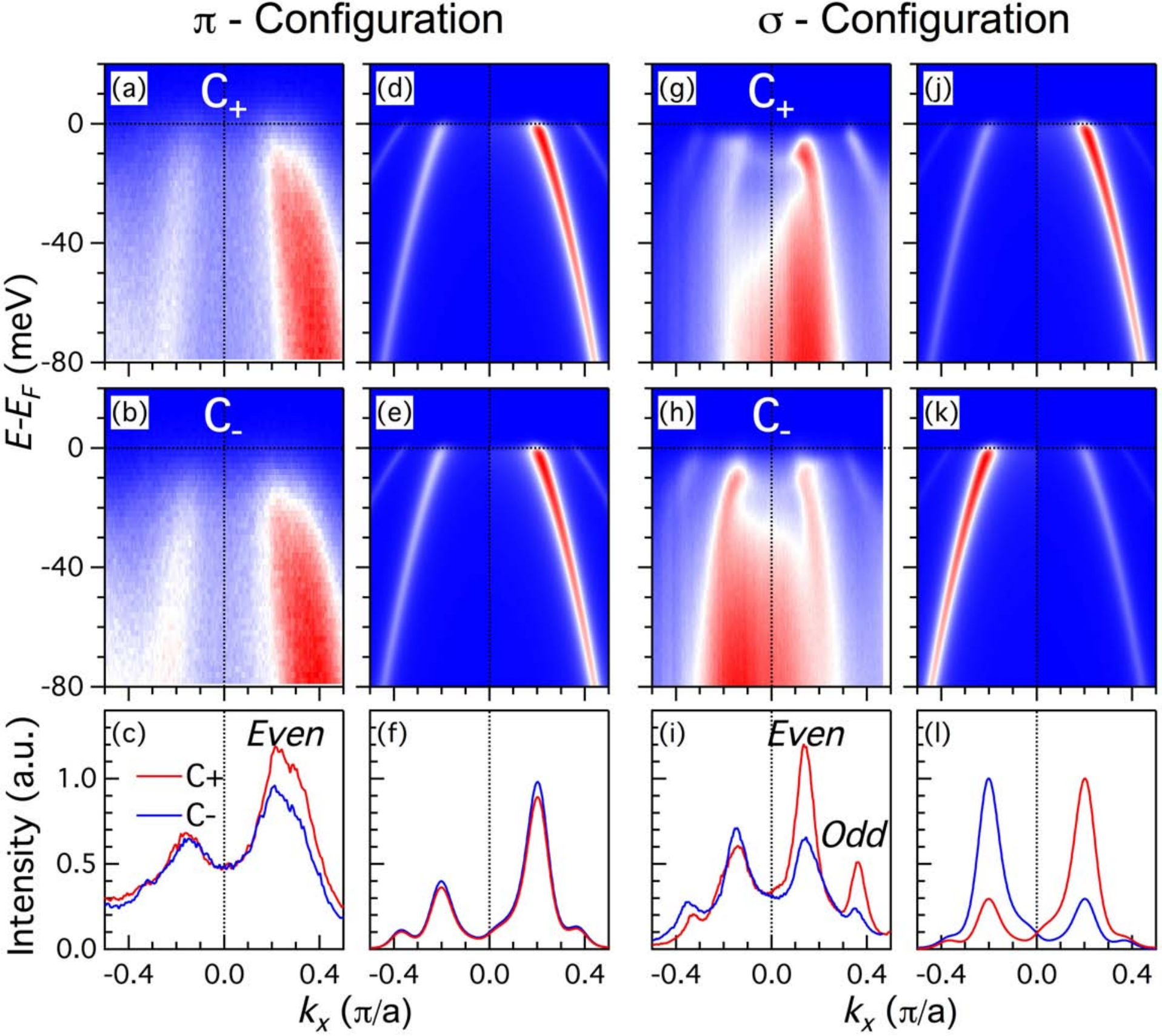}
\end{center}
\caption{\label{Fig_circular}
(Color online) 
(a) and (b) give the ARPES intensity in the $p$ configuration with $C_+$ and $C_-$ polarizations, respectively. (c) MDC profiles near $E_F$ corresponding to (a) and (b). (d) and (e) Simulated ARPES intensity in the $p$ configuration with $C_+$ and $C_-$ polarizations, respectively. (f) MDC profiles near $E_F$ corresponding to (d) and (e). (g) and (h) show the ARPES intensity in the $s$ configuration with $C_+$ and $C_-$ polarizations, respectively. (i) MDC profiles near $E_F$ corresponding to (g) and (h). (j) and (k) Simulated ARPES intensity in the $s$ configuration with $C_+$ and $C_-$ polarizations, respectively. (l) MDC profiles near $E_F$ corresponding to (j) and (k). To simplify the simulations, we chose $\gamma = \pi/4$. 
 }
\end{figure}

We now investigate circular dichroism for the band structure at the $\Gamma$ point and demonstrate that it contains information on the orbital characters of the different bands. Figures \ref{Fig_circular}(a) and \ref{Fig_circular}(b) show the experimental data obtained in the $p$ ARPES configuration using $C_+$ and $C_-$ incoming light, respectively. As expected from the selection rules derived in the previous section for this particular setup and in agreement with our simulations displayed in Figures \ref{Fig_circular}(d) and \ref{Fig_circular}(e), we do not observe strong variations between the two sets of data. This is also confirmed by the near-$E_F$ MDCs shown in Figure \ref{Fig_circular}(c) as well as the simulated ones given in Figure \ref{Fig_circular}(f). Interestingly, the experimental data only show strong intensity for the degenerated inner band [$\alpha$(odd symmetry) and/or $\alpha'$(even symmetry)], but not for the $\beta$(odd symmetry) band. This behavior is captured by our simulations and confirms that the $\beta$ band has a odd symmetry orbital character. From our selection rules, we deduce that mainly the $\alpha'$ band is observed in this configuration.

The situation becomes quite different for the data recorded in the $s$ configuration, once again using $C_+$ and $C_-$ incoming light. The corresponding experimental data are illustrated in Figures \ref{Fig_circular}(g) and \ref{Fig_circular}(h), respectively, and the MDC profiles near $E_F$ are displayed in \ref{Fig_circular}(i). When switching from $C_+$ to $C_-$ polarized light, the observed asymmetry in the intensity is qualitatively reversed with respect to the $\Gamma$ point. As expected from our simulations given in Figures \ref{Fig_circular}(j) and \ref{Fig_circular}(k), the largest switch in the intensity asymmetry is found on the inner band that has an even symmetry orbital characters. Although observed, this effect is less pronounced for the intensity of the $\beta$ band.

\begin{figure}[!t]
\begin{center}
\includegraphics[width=8cm]{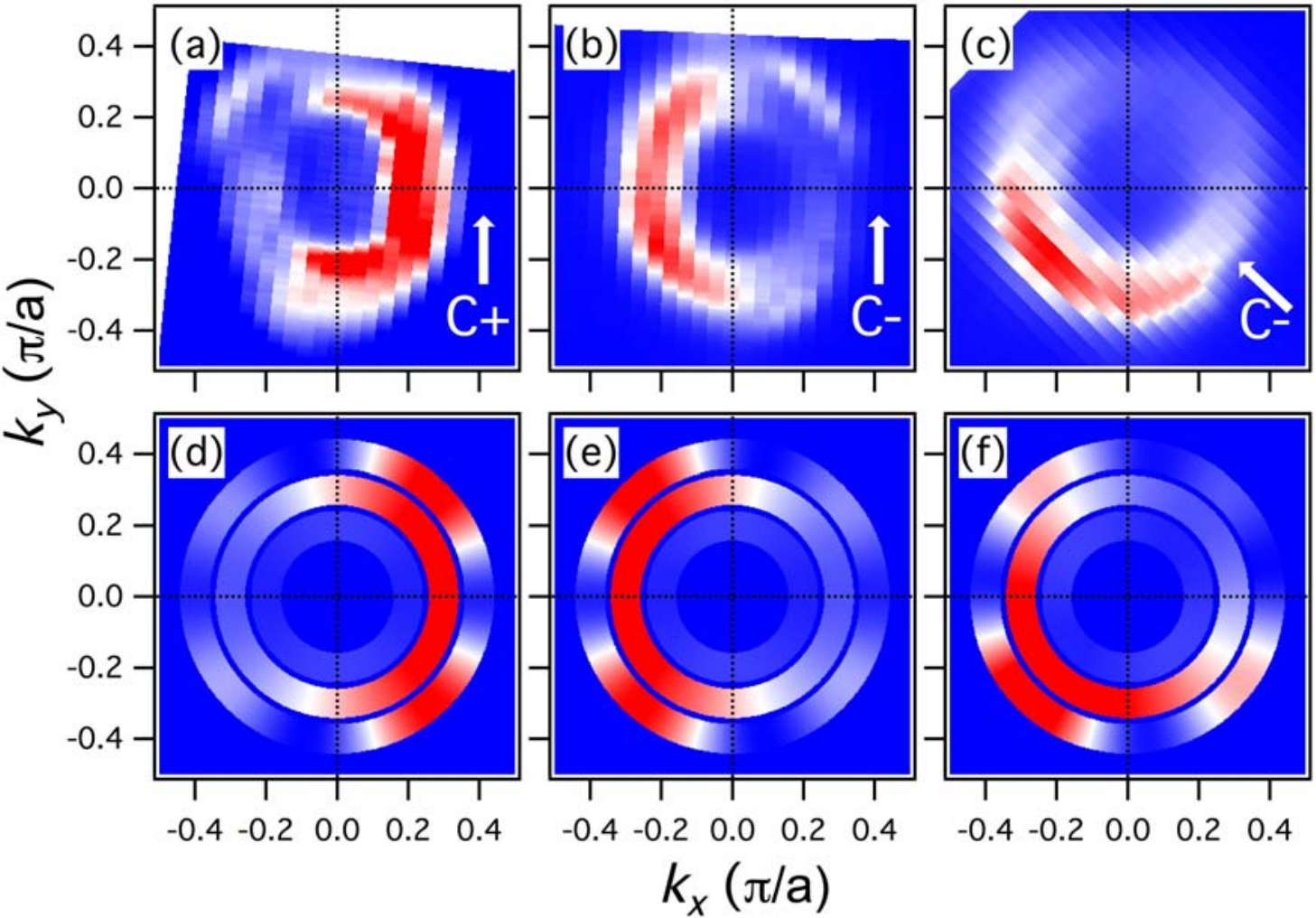}
\end{center}
\caption{\label{circular_FS}
(a)-(c) Fermi surface intensity patterns of Ba$_{0.6}$K$_{0.4}$Fe$_2$As$_2$ recorded with 60 eV circular polarized light. The white arrows show the direction of the incoming light. (d)-(f) Corresponding simulation results.}
\end{figure}

Circular dichroism is also very well illustrated by the Fermi surface intensity patterns recorded on Ba$_{0.6}$K$_{0.4}$Fe$_2$As$_2$ using 60 eV circular polarized light, which are displayed in Figures \ref{circular_FS}(a)-(c). While the intensity on the right side is much stronger when using $C_+$ polarization, the situation is reversed when using $C_-$ light. This effect is also reproduced by our simulations given in Figures \ref{circular_FS}(d)-(f). Interestingly, a comparison of Figures \ref{circular_FS}(b) and (c) indicates that the pattern rotates when the beam incidence rotates as well. It is worth noting that with circular polarized light the minimum of intensity occurs always on one side of the incoming beam direction whereas it is observed away from the incoming beam side when non-polarized light is used, as suggested by Figures \ref{pattern122}(c) and (d).

At this stage we would like to clarify how we determined the phase $\gamma_{\delta}$ that appears in Eq. \eqref{eqMif}. Although we do not understand its complete meaning, which goes beyond the purpose of the current paper, we can intuitively relate this phase to the discontinuity along $k_z$ at the surface of the sample. To fix this parameter, we measured the electronic dispersion along $k_z$, as we now explain. Despite the 3D nature of the crystal and electronic structures of materials measured in ARPES, this technique is so to speak essentially a 2D probe since the momentum perpendicular to the surface exposed is not a good quantum number. However, within the nearly-free electron approximation for the final state \cite{DamascelliPScrypta2004}, access to the third dimension of momentum is often possible by varying the energy of the incident photons. The momentum along the $z$ direction is then given by:

\begin{equation}
k_z = \sqrt{\frac{2m}{\hbar^2}}\sqrt{(h\nu-\Phi-E_B)\cos^2\theta+V_0}
\end{equation}

\noindent where $\theta$ is the angle between the emission direction and the normal to the surface, $m$ is the free electron mass and $V_0$ is to inner potential, which is determined experimentally. 

\begin{figure}[!t]
\begin{center}
\includegraphics[width=9cm]{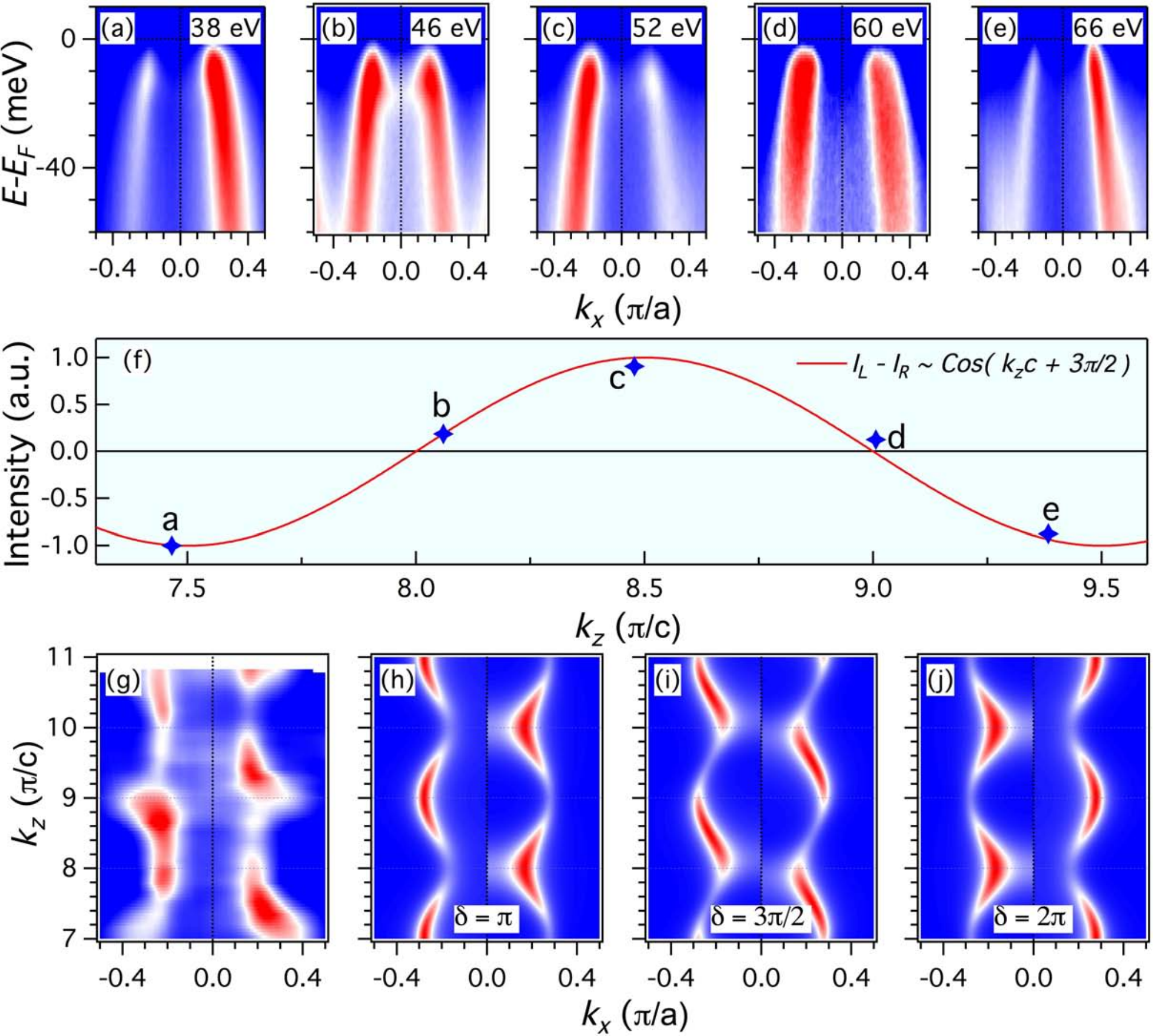}
\end{center}
\caption{\label{Fig_kz}
(Color online) 
(a)-(e) ARPES intensity plots of Ba$_{0.6}$K$_{0.4}$Fe$_2$As$_2$ recorded around the $\Gamma$ point with $C_+$ polarized light. (f) Normalized intensity difference between the left and right sides of panels (a)-(e). The photon energy has been converted into $k_z$ using a inner potential $V_0=14.5$ eV \cite{YM_Xu_NPhys2011}. (g) Photoemission intensity plot of the electronic dispersion of Ba$_{0.6}$K$_{0.4}$Fe$_2$As$_2$ as a function of $k_z$. The data have been recorded between 30 and 90 eV using $C_+$ light. (h)-(j) Simulations corresponding to the experimental conditions in (g). The phase $\gamma_{\delta}$ of Eq. \eqref{eqMif} has been fixed to $k_zc+\delta$.
}
\end{figure}

The photoemission intensity is expected to change with photon energy due to the photoemission cross section \cite{Yeh1985} and can even show resonances at particular photon energies. Photoemission measurements over a wide photon energy range can indeed be used to determine the elemental characters of the states probes \cite{Allen_PRL1990, Richard_PRB2006,QianPRB2011,Ding_J_Phys_Condens_Matter2011}. Experimentally, additional effects that cannot find a simple explanation in the photoemission cross section are observed. Figures \ref{Fig_kz}(a)-(e) show such an interesting phenomenon: the energy-momentum photoemission intensity measured on Ba$_{0.6}$K$_{0.4}$Fe$_2$As$_2$ samples with $C_+$ polarized light exhibits an asymmetry that varies with photon energy. At 38 eV, the left part of the spectrum has a much weaker intensity than the right part. This is no longer the case at 46 eV, where the two sides show almost equivalent intensity. The asymmetry is even reversed at 52 eV, with the left side of the spectrum being much stronger than the right side. The intensity on both becomes almost equal once more at 60 eV before recovering the initial pattern at 66 eV. After finding the $k_z$ correspondence of each photon energy using $V_0=14.5$ eV (similar to the value reported previously \cite{YM_Xu_NPhys2011}), we can plot the normalized intensity difference between the left and right sides of the spectra as a function of $k_z$. The results are displayed in Figure \ref{Fig_kz}(f). Interestingly, the data can be fitted by a cosine function with $k_zc+3\pi/2$ as argument, where $c=6.6$ \AA\xspace is the lattice parameter of the primitive unit cell, which is equivalent to the distance between Fe layers.       

This strange behavior of the photon energy dependence of the intensity extends beyond the 38-66 eV range. Figure \ref{Fig_kz}(g) reveals oscillations in the 30-90 eV range, as $k_z$ goes up and crosses different Brillouin zones. This range corresponds to $k_z$ variations between $7\pi/c$ and $11\pi/c$. The Z positions coincide to $k_z$ values with the largest $k_F$ positions, \emph{i. e.} $k_z = 7\pi/c$, $9\pi/c$ and $11\pi/c$, whereas the $\Gamma$ positions coincide with $k_z = 8\pi/c$ and $10\pi/c$. For each Brillouin zone, the signal on the left-hand side is quite strong as we increase $k_z$ from $\Gamma$ to Z, while the signal is much weaker on the right-hand side. The situation is completely reversed with $k_z$ increasing from Z to $\Gamma$, with the spectral intensity switched from one side to the other. To obtain this effect in the simulations, the phase $\gamma_{\delta}$ has to be fixed to $k_zc+\frac{3\pi}{2}$ over the whole range (see Figure \ref{Fig_kz}(i)). A variation in the phase leads to simulated results completely inconsistent with the experimental data and justify our choice of phase. However, a deeper knowledge of the details of the photoemission process would be needed to provide an \emph{ab initio} value for this parameter.

\section{Orbital characters in $\textrm{FeTe}_{0.55}\textrm{Se}_{0.45}$}

\begin{figure}[!t]
\begin{center}
\includegraphics[width=8cm]{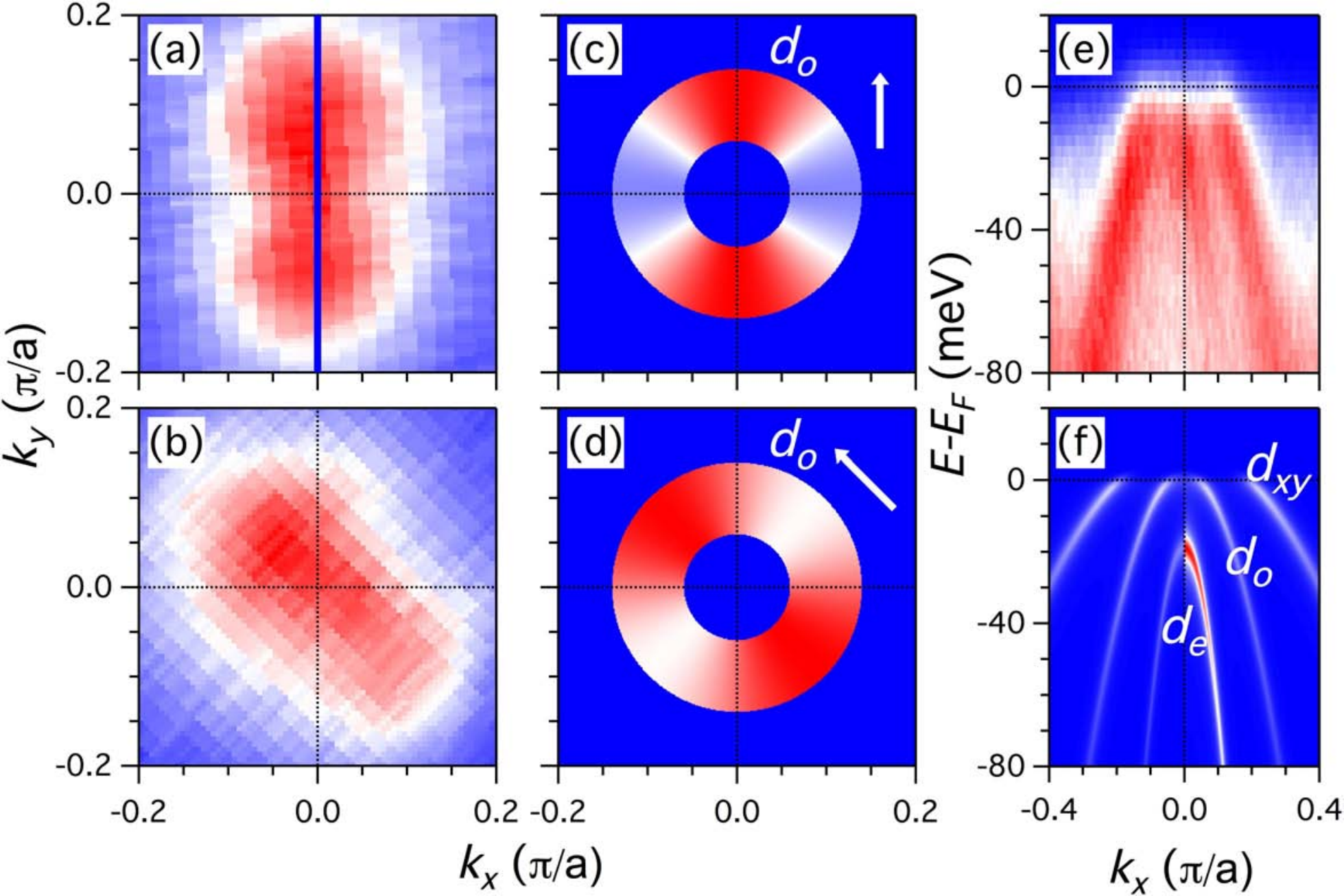}
\end{center}
\caption{\label{FeTeSe}
Fermi surface and band structure of FeTe$_{0.55}$Se$_{0.45}$. (a)-(b) Fermi surface intensity plots recorded with unpolarized photons (21.2 eV) directed along $\Gamma$-M and $\Gamma$-X, respectively. (c)-(d) Corresponding simulations for the $d_o$ band, with the light momentum indicated by the white arrows. (e) ARPES intensity cut recorded along the blue line in (a). (f) Simulations of the intensity cut in (e).}
\end{figure}

We now check our method to determine the orbital characters in FeTe$_{0.55}$Se$_{0.45}$. Figures \ref{FeTeSe}(a) and \ref{FeTeSe}(b) show two Fermi surface intensity patterns of FeTe$_{0.55}$Se$_{0.45}$ recorded with a Helium lamp in the $p$-type ARPES configuration. The two measurements differ only by the orientation of the light momentum, which is aligned along $\Gamma$-M and $\Gamma$-X for Figures \ref{FeTeSe}(a) and \ref{FeTeSe}(b), respectively. In both cases, the Fermi surface patterns exhibit strong two-fold symmetry, with stronger intensity along the light momentum direction. According to our simulations, only the $d_o$ Fermi surface can follow this behavior. The corresponding simulation results for the $d_o$ Fermi surface are given in Figures \ref{FeTeSe}(c) and \ref{FeTeSe}(d), respectively. We note that although small variations in the Fermi surface size do not change qualitatively the simulations, obvious modifications appear when the size is modified significantly. For example, the large $\beta$ Fermi surface in Ba$_{0.6}$K$_{0.4}$Fe$_2$As$_2$ displayed in Figure \ref{pattern122}(d) carries the same dominant orbital character as the much smaller $\beta$ Fermi surface in FeTe$_{0.55}$Se$_{0.45}$ shown in Figure \ref{FeTeSe}(a), which correspond to similar experimental conditions. Yet, both the experimental and theoretical results indicate differences. Nevertheless, in both cases the Fermi surface patterns exhibit a suppression of intensity along the x-axis.

Figure \ref{FeTeSe}(e) shows the ARPES intensity cut along $\Gamma$-M for a light momentum aligned along the same direction (blue line in Figure \ref{FeTeSe}(a)). Two bands are clearly observed, one of them not crossing or barely crossing the Fermi level. Actually, a fine study indicates the presence of the expected third band, which has a much weaker intensity and a $k_F$ only slightly larger than that of the other band crossing the Fermi level \cite{H_Miao}. We display the results of our simulations in Figure \ref{FeTeSe} (f) for a cut in the same configuration, where we assume that the inner band carries a $d_e$ character while the weak outer one is dominated by $d_{xy}$. The main observation is that the $d_e$ band exhibits a strong asymmetry with respect to $\Gamma$. This is indeed what is observed experimentally, reinforcing our assumption. We thus conclude that the outer band has a $d_{xy}$ orbital character.

\section{Discussion}

\begin{figure}[!t]
\begin{center}
\includegraphics[width=8cm]{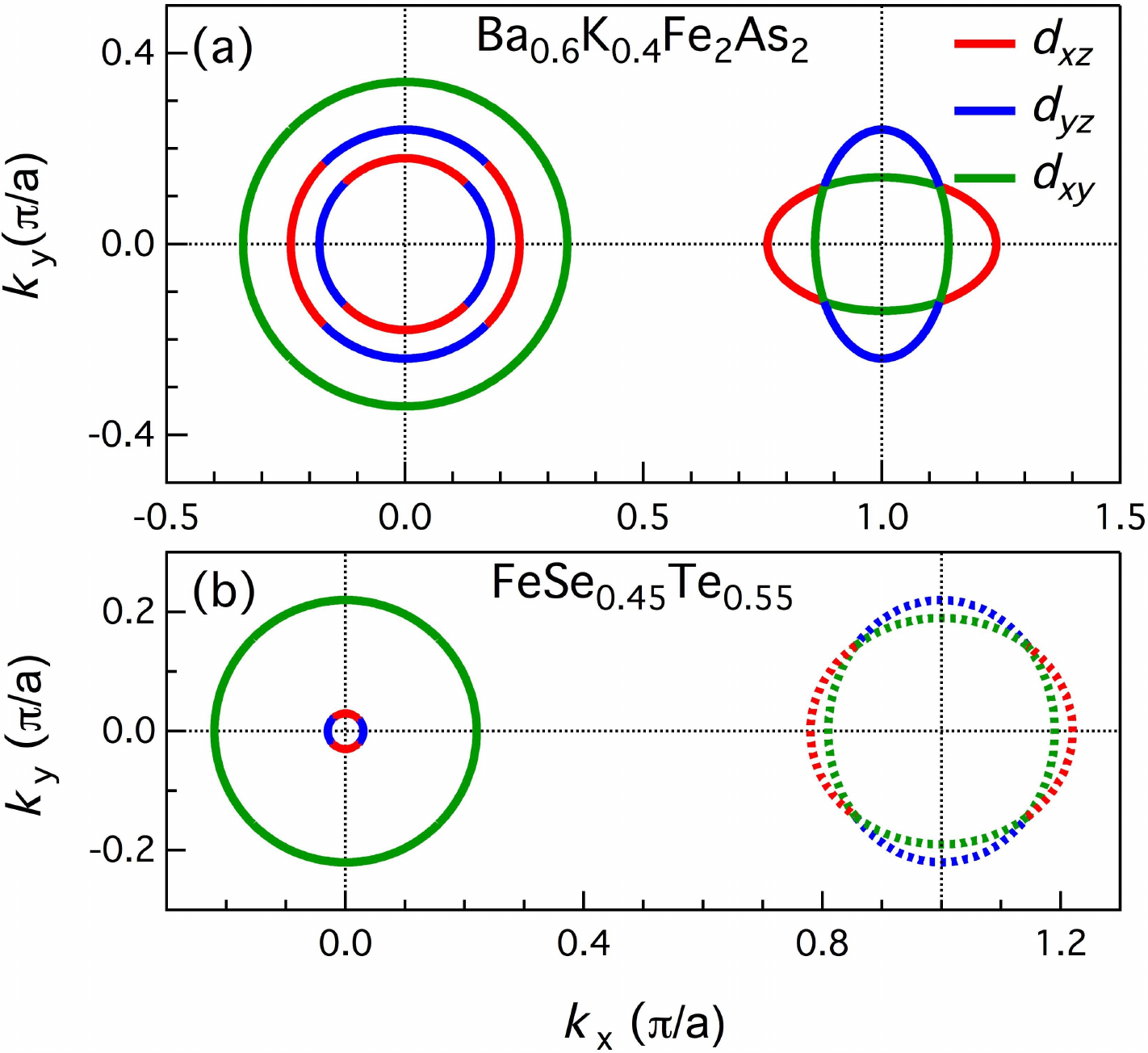}
\end{center}
\caption{\label{summary_orbital}
(Color online) 
Schematic distribution of the orbital characters in Ba$_{0.6}$K$_{0.4}$Fe$_2$As$_2$ and FeTe$_{0.55}$Se$_{0.45}$. Red: $d_{xz}$; Blue: $d_{yz}$; Green: $d_{xy}$.}
\end{figure}

Prior to discuss further the method presented in this paper, we would like to comment on the results obtained for the orbital characterization of the Fermi surface of FeTe$_{0.55}$Se$_{0.45}$ and Ba$_{0.6}$K$_{0.4}$Fe$_2$As$_2$. The summary of our orbital character attributions for the various electronic bands in these materials are displayed in Figure \ref{summary_orbital}. For convenience, we spaced the $\alpha$ and $\alpha'$ Fermi surfaces in Ba$_{0.6}$K$_{0.4}$Fe$_2$As$_2$, which are almost degenerate in the $k_z=0$ plane. Except for absolute and relative variations of the Fermi surface sizes at the $\Gamma$ point, these patterns hold for all $k_z$ values. We stress once more that our experimental observation contrasts with the theoretical expectation of a switch in the orbital distribution of the M-centered Fermi surfaces at $k_z=\pi$ compare to $k_z=0$ \cite{Graser_NJP2009,Graser_PRB2010}, which may have important consequences for inter-pocket interactions \cite{J_KangPRB2011}.

The superconducting gap of Ba$_{0.6}$K$_{0.4}$Fe$_2$As$_2$ is Fermi-surface dependent \cite{Ding_EPL2008,L_ZhaoCPL2008,Nakayama_EPL2009}. More precisely, it is about 12 meV large for all Fermi surface sheets except for the 6 meV gap found on the $\beta$ band, which carries a dominant $d_{xy}$ character. The $2\Delta/k_BT_c$ ratio indicates a pairing in the weak coupling limit for the $\beta$ band. Gaps in the weak coupling regimes are also observed for the $\beta$ band in overdoped Ba$_{0.3}$K$_{0.7}$Fe$_2$As$_2$ \cite{Nakayama_PRB2011} and underdoped Ba$_{0.75}$K$_{0.25}$Fe$_2$As$_2$ \cite{YM_Xu_Ncommun2011}. Interestingly, the 2.5 meV gap size on the $\beta$ band in FeTe$_{0.55}$Se$_{0.45}$ ($T_c=14.5$ K) leads also to a similar ratio \cite{H_Miao}. 

From these observations, one could be tempted to argue that superconducting pairing is controlled by the orbital character, which for some reason could be less efficient for the $d_{xy}$ orbital. However, this argument is in contradiction with the observation of a $d_{xy}$ orbital character at the M point. In reality, the two electronlike ellipses at the M point hybridize and form two distinct Fermi surfaces \cite{Nakayama_EPL2009}. While the inner one is largely dominated by a $d_{xy}$ character, the outer one is formed by a combination of the $d_{xz}$ and $d_{yz}$ orbitals. Both of them show a gap size indicating a strong coupling regime \cite{Nakayama_EPL2009}. A recent study suggests similar results in FeTe$_{0.55}$Se$_{0.45}$ \cite{H_Miao}. Therefore, we conclude that in the Fe-based superconductors there is no direct correlation between the orbital character of a Fermi surface and the gap size. Analyses of the gap size on various Fermi surfaces using gap functions derived from local antiferromagnetic exchange interactions rather suggest that the relative size of the superconducting gap on a particular Fermi surface is determined by its momentum position \cite{YM_Xu_NPhys2011,H_Miao,ZH_LiuPRB2011,JP_Hu2011}.  

The method described in this paper is certainly a reliable and relatively simple way to obtain empirically the orbital characters of bands in the iron-based superconductors. With a proper choice of basis functions, it can be applied to other materials as well. Nevertheless, the model has its own limitations. For example, it remains quite difficult to determine the orbital characters from the Fermi surface patterns in the case of bands with mixed characters. Some theoretical assumptions are often necessary to guide the analysis. For example, we assumed a particular angular distribution for the orbital characters of the electronlike Fermi surfaces forming the Fermi surface at the M point of Fe-based superconductors in order to get a nice agreement between the experimental data and our simulations. However, the method is a powerful tool to discard some scenarios.

Another important limitation concerns the determination of unknown parameters, such as $\gamma_{\delta}$ in Eq. \eqref{eqMif}. As explained in Section \ref{section122}, we imposed the phase $\gamma_{\delta}$ by looking at the photon energy dependence of the Fermi surface pattern. It is clear though that the phase itself may carry some important information that is not accessible directly from our simplify model. From the experimental point of view, further ARPES studies on different materials, involving different electronic orbitals or even different transition metals, may help clarifying this issue.

\section{Summary}

We introduced a simple method to obtain the orbital characters of the various sheets forming the Fermi surface of crystals. The method exploits the asymmetries obtained experimentally in the photoemission intensity patterns of Fermi surface mappings and energy-momentum plots revealed by ARPES in various experimental conditions of beam orientation and light polarization, including non-polarized light. Our method has been successfully applied to Ba$_{0.6}$K$_{0.4}$Fe$_2$As$_2$ and FeTe$_{0.55}$Se$_{0.45}$, which are two Fe-based superconductors. We showed that the multi-sheet Fermi surface of these materials originates mainly from Fe $3d$ electrons with $d_{xy}$, $d_{xz}$ and $d_{yz}$ orbital characters. Our results suggest that there is no direct relationship between the strength of the superconducting gap on the various Fermi surface sheets of these multi-band systems and the orbital characters from which they are mainly formed.  


\begin{acknowledgments}
We acknowledge T. Kim, E. Rienks, E. Razzoli for technical assistance, and Z. Fang, Z. Wang, S.-J. Jiang  for useful discussions. We are also grateful to G.-F. Chen, N.-L. Wang and G.-D. Gu who provided the high-quality single crystals necessary for this study. This work is supported by the Chinese Academy of Sciences (grant No. 2010Y1JB6), the Ministry of Science and Technology of China (grants No. 2010CB923000, No. 2011CBA0010), and the Nature Science Foundation of China (grants No. 10974175, No. 11004232, and No. 11050110422) and the Sino-Swiss Science and Technology Cooperation (2010). This work was partly performed at the Swiss Light Source, Paul Scherrer Institut, Villigen, Switzerland, and at BESSY, Helmholtz Zentrum, Berlin, Germany.
\end{acknowledgments}

\onecolumngrid
\appendix
\section{Details of the matrix element calculations}
In the following, we define: 
\begin{eqnarray}
	\braket{\vec{r}}{f}=e^{i\vec{k_f}\cdot \vec{r}}&=&4\pi\sum_{l=0}^{\infty}i^lj_l(k_fr)\sum_{m=-l}^lY_l^{m*}(\theta_k, \phi_k) Y_l^m(\theta,\phi)\\
	\braket{\vec{r}}{nl_0m_0}&=&R_{nl_0}(r)Y_{l_0}^{m_0}(\theta,\phi)\\
	\vec{A}\cdot\vec{r}&=&\sum_{\alpha=1}^{3}A_\alpha r_\alpha\\
	r_\alpha&=&r\sqrt{\frac{4\pi}{3}}p_\alpha	
\end{eqnarray}
where, $\alpha=x,y,z. p_x,p_y,p_z$ are the $p$-orbital angular distribution functions. Within the 3-step model approximation, the photoemission matrix element between an initial state $\ket{nl_0m_0}$ and a final state $\ket{f}$ becomes

\begin{eqnarray}
\matel{f}{\vec{A}\cdot\vec{r}}{nl_0m_0}&=&\sum_{\alpha=1}^{3}A_\alpha\sum_{l=0}^{\infty}i^l\cdot4\pi\int_{0}^{\infty}drj_l(k_fr)R_{nl_0}(r)r^3\cdot\sum_{m=-l}^{l}Y_l^{m*}(\theta_k,\phi_k)\cdot\oint d\Omega Y_l^{m}p_\alpha Y_{l_0}^{m_0}\\
&=&\sum_{\alpha=1}^{3}A_\alpha\sum_{l=0}^{\infty}i^l\cdot\rho_l^{nl_0}(k_f)\cdot\sum_{m=-l}^{l}Y_l^{m*}(\theta_k,\phi_k)\cdot\oint d\Omega Y_l^{m}p_\alpha Y_{l_0}^{m_0}\\
&=&\sum_{\alpha=1}^{3}A_\alpha\sum_{m_1=-1}^{1}\alpha_{m_1}\sum_{l=0}^{\infty}i^l\cdot\rho_l^{nl_0}(k_f)\cdot\sum_{m=-l}^{l}Y_l^{m*}(\theta_k,\phi_k)\cdot\oint d\Omega Y_l^{m}Y_{1}^{m_1}Y_{l_0}^{m_0}\label{appendix7}
\end{eqnarray}

\noindent Where we defined $p_\alpha\equiv\sum_{m_1=-1}^{1}\alpha_{m_1}Y_{1}^{m_1}$.
\\
\\
We then use Wigner's formalism for the 3j-symbols:
\begin{eqnarray}
\oint d\Omega Y_l^{m}Y_{1}^{m_1}Y_{l_0}^{m_0} = \sqrt{\frac{(2l+1)\cdot3\cdot(2l_0+1)}{4\pi}}
(\begin{matrix}
l & 1 & l_0 \\
0 &0 & 0
\end{matrix})
(\begin{matrix}
l & 1 & l_0 \\
m &m_1 & m_0
\end{matrix})\equiv W(\begin{matrix}
l & 1 & l_0 \\
m &m_1 & m_0
\end{matrix})
\end{eqnarray}
which may be non-zero only if $l_0\neq0$, $m=-(m_1+m_0)$ and $l=l_0-1\textrm{ or } l_0+1$. \\
Hence,
\begin{eqnarray}
\oint d\Omega Y_l^{m}Y_{1}^{m_1}Y_{l_0}^{m_0} = \delta_{l,l_0-1}^{m,-(m_0+m_1)}W(\begin{matrix}
l & 1 & l_0 \\
m &m_1 & m_0
\end{matrix})
+
\delta_{l,l_0+1}^{m,-(m_0+m_1)}W(\begin{matrix}
l & 1 & l_0 \\
m &m_1 & m_0
\end{matrix})
\end{eqnarray}

\noindent Substituting the previous result into equation \eqref{appendix7}, we get:
\begin{eqnarray}
\matel{f}{\vec{A}\cdot\vec{r}}{nl_0m_0}&=&\sum_{i=\alpha}^{3}A_\alpha\sum_{m_1=-1}^{1}\alpha_{m_1}\\
&\cdot& [i^{l_0-1}\rho_{l_0-1}^{nl_0}(k_f)(-1)^{m_0+m_1}Y_{l_0-1}^{m_0+m_1}(\theta_k,\phi_k)W\left(\begin{matrix}l_0-1 & 1 & l_0 \\ -(m_0+m_1) & m_1 & m_0 \end{matrix}\right) \\
&+&i^{l_0+1}\rho_{l_0+1}^{nl_0}(k_f)(-1)^{m_0+m_1}Y_{l_0+1}^{m_0+m_1}(\theta_k,\phi_k)W\left(\begin{matrix}l_0+1 & 1 & l_0 \\ -(m_0+m_1) & m_1 & m_0 \end{matrix}\right) ]\\
&\equiv&\sum_{\alpha=1}^{3}A_\alpha\sum_{m_1=-1}^{1}\alpha_{m_1}(g_-^n(m_1,l_0,m_0,\vec{k})+g_+^n(m_1,l_0,m_0,\vec{k}))\\
&\equiv&\sum_{\alpha=1}^{3}A_\alpha\sum_{m_1=-1}^{1}\alpha_{m_1}G_n(m_1,l_0,m_0,\vec{k})\\
&\equiv&\sum_{\alpha=1}^{3}A_\alpha\cdot M_n(\alpha,l_0,m_0,\vec{k})
\end{eqnarray}
Where, $\vec{k}=(k_f,\theta_k, \phi_k)$ and
\begin{eqnarray}
M_x(l_0,m_0,\vec{k}) &=& M_n(1,l_0,m_0,\vec{k}) = \sqrt{\frac{1}{2}}(G_n(-1,l_0,m_0,\vec{k})-G_n(1,l_0,m_0,\vec{k}))\nonumber\\
M_y(l_0,m_0,\vec{k})  &=& M_n(2,l_0,m_0,\vec{k})= i\sqrt{\frac{1}{2}}(G_n(-1,l_0,m_0,\vec{k})+G_n(1,l_0,m_0,\vec{k}))\nonumber\\
M_z(l_0,m_0,\vec{k})  &=& M_n(3,l_0,m_0,\vec{k}) = G_n(0,l_0,m_0,\vec{k})
\end{eqnarray}

\noindent We now apply above results to the case of $3d$ electrons, for which $l_0=2$. We obtain:
\begin{eqnarray}
 M_{\alpha}^{d_{z^2}}(\vec{k})  &=& M_\alpha(2,0,\vec{k}) \nonumber
 \\
 M_{\alpha}^{d_{yz}}(\vec{k})  &=& i\sqrt{\frac{1}{2}}(M_\alpha(2,-1,\vec{k})+M_\alpha(2,1,\vec{k}))\nonumber
\\
 M_{\alpha}^{d_{xz}}(\vec{k})  &=& \sqrt{\frac{1}{2}}(M_\alpha(2,-1,\vec{k})-M_\alpha(2,1,\vec{k}))\nonumber
\\
 M_{\alpha}^{d_{xy}}(\vec{k})  &=&  i\sqrt{\frac{1}{2}}(M_\alpha(2,-2,\vec{k})-M_\alpha(2,2,\vec{k}))\nonumber
\\
 M_{\alpha}^{d_{x^2-y^2}}(\vec{k})  &=&  \sqrt{\frac{1}{2}}(M_\alpha(2,-2,\vec{k})+M_\alpha(2,2,\vec{k}))
 \end{eqnarray}
 
\noindent The even ($d_e$) and odd ($d_o$) combinations of these matrices for the $d_{xz}$ and $d_{yz}$ orbitals around the $\Gamma$ point are ($\theta_{FS}=\phi_k$):
 \begin{eqnarray}
 M_{\alpha}^{d_e}(\vec{k})  &=& \cos\phi_k M_{\alpha}^{d_{xz}}(\vec{k})+\sin\phi_k M_{\alpha}^{d_{yz}}(\vec{k})  \nonumber\\
  M_{\alpha}^{d_o}(\vec{k})  &=& -\sin\phi_k M_{\alpha}^{d_{xz}}(\vec{k})+\cos\phi_k M_{\alpha}^{d_{yz}}(\vec{k})  
 \end{eqnarray}

\noindent We can express the previous results by defining:
\begin{eqnarray}
Y_l^m(\theta,\phi) = C_l^mP_l^m(\theta)e^{im\phi}
\end{eqnarray}
where $P_l^m(\theta)$ contains all the $\theta$ dependence and $C_l^m$ contains all the numerical prefactors. For example, we have:
\begin{eqnarray}
Y_3^1(\theta,\phi) &=& \frac{-1}{8}\sqrt{\frac{21}{\pi}}\cdot \sin\theta(5\cos^2\theta-1)\cdot e^{i\phi}\nonumber\\
&=& C_3^1\cdot P_3^1(\theta)\cdot e^{i\phi}
\end{eqnarray}

\noindent For $3d$ electrons, this leads to equations \eqref{appendix_Mz2}-\eqref{appendix_Mx2_y2}:\\
\begin{eqnarray}
&d_{z^2}&\nonumber\\
M_x(d_{z^2}) &=& -i\sqrt{2}C_1^1C_2^0\left[\frac{2}{5}\rho_1(k_f)P_1^1+\frac{3}{5}\rho_3(k_f)P_3^1\right]\{-\cos\phi_k\}\nonumber\\
M_y(d_{z^2}) &=& -i\sqrt{2}C_1^1C_2^0\left[\frac{2}{5}\rho_1(k_f)P_1^1+\frac{3}{5}\rho_3(k_f)P_3^1\right]\{-\sin\phi_k\}\nonumber\\
M_z(d_{z^2}) &=& iC_1^0C_2^0\left[-\frac{4}{5}\rho_1(k_f)P_1^0+\frac{3}{5}\rho_3(k_f)P_3^0\right]\{-1\}\label{appendix_Mz2}
 \end{eqnarray}
 
 \begin{eqnarray}
&d_{xz}&\nonumber\\
M_x(d_{xz}) &=& iC_1^1C_2^1\left[\frac{2}{5}\rho_1(k_f)P_1^0+\frac{1}{5}\rho_3(k_f)P_3^0-\rho_3(k_f)P_3^2\cdot \cos 2\phi_k\right]\{1\}\nonumber\\
M_y(d_{xz}) &=& iC_1^1C_2^1\left[-\rho_3(k_f)P_3^2\right]\sin 2\phi_k\sim\{0\}\nonumber\\
M_z(d_{xz}) &=& -i\sqrt{2}C_1^0C_2^1\left[-\frac{1}{5}\rho_1(k_f)P_1^1+\frac{1}{5}\rho_3(k_f)P_3^1\right]\{-\cos\phi_k\}
 \end{eqnarray}

 \begin{eqnarray}
&d_{yz}&\nonumber\\
M_x(d_{yz}) &=&  iC_1^1C_2^1\left[-\rho_3(k_f)P_3^2\right]\sin 2\phi_k \sim\{0\}\nonumber\\
M_y(d_{yz}) &=& iC_1^1C_2^1\left[\frac{2}{5}\rho_1(k_f)P_1^0+\frac{1}{5}\rho_3(k_f)P_3^0+\rho_3(k_f)P_3^2\cdot \cos2\phi_k\right]\{1\}\nonumber\\
M_z(d_{yz}) &=& -i\sqrt{2}C_1^0C_2^1\left[-\frac{1}{5}\rho_1(k_f)P_1^1+\frac{1}{5}\rho_3(k_f)P_3^1\right]\{-\sin\phi_k\}
 \end{eqnarray}

  \begin{eqnarray}
&d_{xy}&\nonumber\\
M_x(d_{xy}) &=& -iC_1^1C_2^2\left[\frac{4}{5}\rho_1(k_f)P_1^1+\frac{1}{5}\rho_3(k_f)P_3^1-\rho_3(k_f)P_3^3(4\cos^2\phi_k-1)\right]\{\sin\phi_k\}\nonumber\\
M_y(d_{xy}) &=& -iC_1^1C_2^2\left[\frac{4}{5}\rho_1(k_f)P_1^1+\frac{1}{5}\rho_3(k_f)P_3^1+\rho_3(k_f)P_3^3(4\cos^2\phi_k-3)\right]\{\cos\phi_k\}\nonumber\\
M_z(d_{xy}) &=& i\sqrt{2}C_1^0C_2^2\left[\rho_3(k_f)P_3^2\right]\{ -\sin2\phi_k\}
 \end{eqnarray}

  \begin{eqnarray}
&d_{x^2-y^2}&\nonumber\\
M_x(d_{x^2-y^2}) &=& -iC_1^1C_2^2\left[\frac{4}{5}\rho_1(k_f)P_1^1+\frac{1}{5}\rho_3(k_f)P_3^1-\rho_3(k_f)P_3^3(4cos^2\phi_k-3)\right]\{\cos\phi_k\}\nonumber\\
M_y(d_{x^2-y^2}) &=&  -iC_1^1C_2^2\left[\frac{4}{5}\rho_1(k_f)P_1^1+\frac{1}{5}\rho_3(k_f)P_3^1+\rho_3(k_f)P_3^3(4cos^2\phi_k-1)\right]\{-\sin\phi_k\}\nonumber\\
M_z(d_{x^2-y^2}) &=& i\sqrt{2}C_1^0C_2^2\left[\rho_3(k_f)P_3^2\right]\{-\cos2\phi_k\}\label{appendix_Mx2_y2}
 \end{eqnarray}
 
 The main idea behind our simplified approach is to neglect the $\phi_k$ dependence of the prefactors preceding the curly brackets in the previous equations. We note that $M_x(d_{yz})$ and $M_y(d_{xz})$ are set to zero since the terms preceding the $\sin 2\phi_k$ function are vanishingly small compared to the other matrix components. The curly bracket terms correspond to the components of the simplified matrices given in Section \ref{computational_details}.

\twocolumngrid
\bibliography{Wangref}

\begin{thebibliography}{44}%
\makeatletter
\providecommand \@ifxundefined [1]{%
 \@ifx{#1\undefined}
}%
\providecommand \@ifnum [1]{%
 \ifnum #1\expandafter \@firstoftwo
 \else \expandafter \@secondoftwo
 \fi
}%
\providecommand \@ifx [1]{%
 \ifx #1\expandafter \@firstoftwo
 \else \expandafter \@secondoftwo
 \fi
}%
\providecommand \natexlab [1]{#1}%
\providecommand \enquote  [1]{``#1''}%
\providecommand \bibnamefont  [1]{#1}%
\providecommand \bibfnamefont [1]{#1}%
\providecommand \citenamefont [1]{#1}%
\providecommand \href@noop [0]{\@secondoftwo}%
\providecommand \href [0]{\begingroup \@sanitize@url \@href}%
\providecommand \@href[1]{\@@startlink{#1}\@@href}%
\providecommand \@@href[1]{\endgroup#1\@@endlink}%
\providecommand \@sanitize@url [0]{\catcode `\\12\catcode `\$12\catcode
  `\&12\catcode `\#12\catcode `\^12\catcode `\_12\catcode `\%12\relax}%
\providecommand \@@startlink[1]{}%
\providecommand \@@endlink[0]{}%
\providecommand \url  [0]{\begingroup\@sanitize@url \@url }%
\providecommand \@url [1]{\endgroup\@href {#1}{\urlprefix }}%
\providecommand \urlprefix  [0]{URL }%
\providecommand \Eprint [0]{\href }%
\@ifxundefined \urlstyle {%
  \providecommand \doi  [0]{\begingroup \@sanitize@url \@doi}%
  \providecommand \@doi [1]{\endgroup \@@startlink {\doibase
  #1}doi:\discretionary {}{}{}#1\@@endlink }%
}{%
  \providecommand \doi  [0]{doi:\discretionary{}{}{}\begingroup
  \urlstyle{rm}\Url }%
}%
\providecommand \doibase [0]{http://dx.doi.org/}%
\providecommand \Doi [0]{\begingroup \@sanitize@url \@Doi }%
\providecommand \@Doi  [1]{\endgroup\@@startlink{\doibase#1}\@@Doi}%
\providecommand \@@Doi [1]{#1\@@endlink}%
\providecommand \selectlanguage [0]{\@gobble}%
\providecommand \bibinfo  [0]{\@secondoftwo}%
\providecommand \bibfield  [0]{\@secondoftwo}%
\providecommand \translation [1]{[#1]}%
\providecommand \BibitemOpen [0]{}%
\providecommand \bibitemStop [0]{}%
\providecommand \bibitemNoStop [0]{.\EOS\space}%
\providecommand \EOS [0]{\spacefactor3000\relax}%
\providecommand \BibitemShut  [1]{\csname bibitem#1\endcsname}%
\bibitem [{\citenamefont {{J. W. Allen, C. G. Olson, M. B. Maple, J.-S. Kang,
  L. Z. Liu, J.-H. Park, R. O. Anderson, W. P. Ellis, J. T. Markert, Y.
  Dalichaouch and R. Liu}}(1990)}]{Allen_PRL1990}%
  \BibitemOpen
  \bibfield  {author} {\bibinfo {author} {\bibnamefont {{J. W. Allen, C. G.
  Olson, M. B. Maple, J.-S. Kang, L. Z. Liu, J.-H. Park, R. O. Anderson, W. P.
  Ellis, J. T. Markert, Y. Dalichaouch and R. Liu}}},\ }\href@noop {}
  {\bibfield  {journal} {\bibinfo  {journal} {Phys. Rev. Lett.},\ }\textbf
  {\bibinfo {volume} {64}},\ \bibinfo {pages} {595} (\bibinfo {year}
  {1990})}\BibitemShut {NoStop}%
\bibitem [{\citenamefont {{F. Ronning, K. M. Shen, N. P. Armitage, A.
  Damascelli, D. H. Lu, Z.-X. Shen, L. L. Miller and C.
  Kim}}(2005)}]{Ronning_PRB2005}%
  \BibitemOpen
  \bibfield  {author} {\bibinfo {author} {\bibnamefont {{F. Ronning, K. M.
  Shen, N. P. Armitage, A. Damascelli, D. H. Lu, Z.-X. Shen, L. L. Miller and
  C. Kim}}},\ }\href@noop {} {\bibfield  {journal} {\bibinfo  {journal} {Phys.
  Rev. B},\ }\textbf {\bibinfo {volume} {71}},\ \bibinfo {pages} {094518}
  (\bibinfo {year} {2005})}\BibitemShut {NoStop}%
\bibitem [{\citenamefont {{P. Richard, Z.-H. Pan, M. Neupane, A. V. Fedorov, T.
  Valla, P. D. Johnson, G. D. Gu, W. Ku, Z. Wang and H.
  Ding}}(2006)}]{Richard_PRB2006}%
  \BibitemOpen
  \bibfield  {author} {\bibinfo {author} {\bibnamefont {{P. Richard, Z.-H. Pan,
  M. Neupane, A. V. Fedorov, T. Valla, P. D. Johnson, G. D. Gu, W. Ku, Z. Wang
  and H. Ding}}},\ }\href@noop {} {\bibfield  {journal} {\bibinfo  {journal}
  {Phys. Rev. B},\ }\textbf {\bibinfo {volume} {74}},\ \bibinfo {pages}
  {094512} (\bibinfo {year} {2006})}\BibitemShut {NoStop}%
\bibitem [{\citenamefont {{A. F. Santander-Syro \emph{et
  al.}}}(2010)}]{Santander_Nature2010}%
  \BibitemOpen
  \bibfield  {author} {\bibinfo {author} {\bibnamefont {{A. F. Santander-Syro
  \emph{et al.}}}},\ }\href@noop {} {\bibfield  {journal} {\bibinfo  {journal}
  {Nature},\ }\textbf {\bibinfo {volume} {469}},\ \bibinfo {pages} {189}
  (\bibinfo {year} {2010})}\BibitemShut {NoStop}%
\bibitem [{\citenamefont {{T. Qian, N. Xu, Y.-B. Shi, K. Nakayama, P. Richard,
  T. Kawahara, T. Sato, T. Takahashi, M. Neupane, Y.-M. Xu, X.-P. Wang, G. Xu,
  X. Dai, Z. Fang, P. Cheng, H.-H. Wen and H. Ding}}(2011)}]{QianPRB2011}%
  \BibitemOpen
  \bibfield  {author} {\bibinfo {author} {\bibnamefont {{T. Qian, N. Xu, Y.-B.
  Shi, K. Nakayama, P. Richard, T. Kawahara, T. Sato, T. Takahashi, M. Neupane,
  Y.-M. Xu, X.-P. Wang, G. Xu, X. Dai, Z. Fang, P. Cheng, H.-H. Wen and H.
  Ding}}},\ }\href@noop {} {\bibfield  {journal} {\bibinfo  {journal} {Phys.
  Rev. B},\ }\textbf {\bibinfo {volume} {83}},\ \bibinfo {pages} {140513(R)}
  (\bibinfo {year} {2011})}\BibitemShut {NoStop}%
\bibitem [{\citenamefont {{H. Ding, K. Nakayama, P. Richard, S. Souma, T. Sato,
  T. Takahashi, M. Neupane, Y.-M. Xu, Z.-H. Pan, A.V. Fedorov, Z.Wang, X. Dai,
  Z. Fang, G. F. Chen, J. L. Luo and N. L.
  Wang}}(2011)}]{Ding_J_Phys_Condens_Matter2011}%
  \BibitemOpen
  \bibfield  {author} {\bibinfo {author} {\bibnamefont {{H. Ding, K. Nakayama,
  P. Richard, S. Souma, T. Sato, T. Takahashi, M. Neupane, Y.-M. Xu, Z.-H. Pan,
  A.V. Fedorov, Z.Wang, X. Dai, Z. Fang, G. F. Chen, J. L. Luo and N. L.
  Wang}}},\ }\href@noop {} {\bibfield  {journal} {\bibinfo  {journal} {J. Phys:
  Condens. Matter},\ }\textbf {\bibinfo {volume} {23}},\ \bibinfo {pages}
  {135501} (\bibinfo {year} {2011})}\BibitemShut {NoStop}%
\bibitem [{\citenamefont {{C. Hwang, C.-H. Park, D. A. Siegel, A. V. Fedorov,
  S. G. Louie and A. Lanzara}}(2011)}]{Hwang_PRB2011}%
  \BibitemOpen
  \bibfield  {author} {\bibinfo {author} {\bibnamefont {{C. Hwang, C.-H. Park,
  D. A. Siegel, A. V. Fedorov, S. G. Louie and A. Lanzara}}},\ }\href@noop {}
  {\bibfield  {journal} {\bibinfo  {journal} {Phys. Rev. B},\ }\textbf
  {\bibinfo {volume} {84}},\ \bibinfo {pages} {125422} (\bibinfo {year}
  {2011})}\BibitemShut {NoStop}%
\bibitem [{\citenamefont {{A. Bansil, M. Lindroos, S. Sahrakorpi and R. S.
  Markiewicz}}(2005)}]{Bansil_PRB2005}%
  \BibitemOpen
  \bibfield  {author} {\bibinfo {author} {\bibnamefont {{A. Bansil, M.
  Lindroos, S. Sahrakorpi and R. S. Markiewicz}}},\ }\href@noop {} {\bibfield
  {journal} {\bibinfo  {journal} {Phys. Rev. B},\ }\textbf {\bibinfo {volume}
  {71}},\ \bibinfo {pages} {012503} (\bibinfo {year} {2005})}\BibitemShut
  {NoStop}%
\bibitem [{\citenamefont {{L. Roca, M. Izquierdo, A. Tejeda, G.D. Gu, J. Avila,
  M.C. Asensio}}(2003)}]{Roca_2003}%
  \BibitemOpen
  \bibfield  {author} {\bibinfo {author} {\bibnamefont {{L. Roca, M. Izquierdo,
  A. Tejeda, G.D. Gu, J. Avila, M.C. Asensio}}},\ }\href@noop {} {\bibfield
  {journal} {\bibinfo  {journal} {Applied Surface Science},\ }\textbf {\bibinfo
  {volume} {212-213}},\ \bibinfo {pages} {62} (\bibinfo {year}
  {2003})}\BibitemShut {NoStop}%
\bibitem [{\citenamefont {{M. Mulazzi, M. Hochstrasser, M. Corso, I. Vobornik,
  J. Fujii, J. Osterwalder, J. Henk and G. Rossi}}(2006)}]{Mulazzi_PRB2006}%
  \BibitemOpen
  \bibfield  {author} {\bibinfo {author} {\bibnamefont {{M. Mulazzi, M.
  Hochstrasser, M. Corso, I. Vobornik, J. Fujii, J. Osterwalder, J. Henk and G.
  Rossi}}},\ }\href@noop {} {\bibfield  {journal} {\bibinfo  {journal} {Phys.
  Rev. B},\ }\textbf {\bibinfo {volume} {74}},\ \bibinfo {pages} {035118}
  (\bibinfo {year} {2006})}\BibitemShut {NoStop}%
\bibitem [{\citenamefont {{P Richard, T Sato, K Nakayama, T. Takahashi and H.
  Ding}}(2011)}]{Richard_PoPP2011}%
  \BibitemOpen
  \bibfield  {author} {\bibinfo {author} {\bibnamefont {{P Richard, T Sato, K
  Nakayama, T. Takahashi and H. Ding}}},\ }\href@noop {} {\bibfield  {journal}
  {\bibinfo  {journal} {Rep. Prog. Phys.},\ }\textbf {\bibinfo {volume} {74}},\
  \bibinfo {pages} {124512} (\bibinfo {year} {2011})}\BibitemShut {NoStop}%
\bibitem [{\citenamefont {{H. Kontani and S. Onari}}(2010)}]{KontaniPRL2010}%
  \BibitemOpen
  \bibfield  {author} {\bibinfo {author} {\bibnamefont {{H. Kontani and S.
  Onari}}},\ }\href@noop {} {\bibfield  {journal} {\bibinfo  {journal} {Phys.
  Rev. Lett.},\ }\textbf {\bibinfo {volume} {104}},\ \bibinfo {pages} {157001}
  (\bibinfo {year} {2010})}\BibitemShut {NoStop}%
\bibitem [{\citenamefont {{S. H\"{u}fner}}(1995)}]{Hufner_Photoemission}%
  \BibitemOpen
  \bibfield  {author} {\bibinfo {author} {\bibnamefont {{S. H\"{u}fner}}},\
  }\href@noop {} {\bibfield  {journal} {\bibinfo  {journal} {\emph{Photoemssion
  Spectroscopy, Principle and Applications, 2$^{nd}$ Ed.}, Springer-Verlag,
  Berlin, Germany}} (\bibinfo {year} {1995})}\BibitemShut {NoStop}%
\bibitem [{\citenamefont {{J. Fink, S. Thirupathaiah, R. Ovsyannikov, H. A.
  D\"{u}rr, R. Follath, Y. Huang, S. de Jong, M. S. Golden, Y.-Z. Zhang, H. O.
  Jeschke, R. Valent\'{i}, C. Felser, S. Dastjani Farahani, M. Rotter and D.
  Johrendt}}(2009)}]{FinkPRB2009}%
  \BibitemOpen
  \bibfield  {author} {\bibinfo {author} {\bibnamefont {{J. Fink, S.
  Thirupathaiah, R. Ovsyannikov, H. A. D\"{u}rr, R. Follath, Y. Huang, S. de
  Jong, M. S. Golden, Y.-Z. Zhang, H. O. Jeschke, R. Valent\'{i}, C. Felser, S.
  Dastjani Farahani, M. Rotter and D. Johrendt}}},\ }\href@noop {} {\bibfield
  {journal} {\bibinfo  {journal} {Phys. Rev. B},\ }\textbf {\bibinfo {volume}
  {79}},\ \bibinfo {pages} {155118} (\bibinfo {year} {2009})}\BibitemShut
  {NoStop}%
\bibitem [{\citenamefont {{W. Malaeb, T. Yoshida, A. Fujimori, M. Kubota, K.
  Ono, K. Kihou, P. M. Shirage, H. Kito, A. Iyo, H. Eisaki, Y. Nakajima, T.
  Tamegai and R. Arita}}(2009)}]{MalaebJPSJ2009}%
  \BibitemOpen
  \bibfield  {author} {\bibinfo {author} {\bibnamefont {{W. Malaeb, T. Yoshida,
  A. Fujimori, M. Kubota, K. Ono, K. Kihou, P. M. Shirage, H. Kito, A. Iyo, H.
  Eisaki, Y. Nakajima, T. Tamegai and R. Arita}}},\ }\href@noop {} {\bibfield
  {journal} {\bibinfo  {journal} {J. Phys. Soc. Jpn.},\ }\textbf {\bibinfo
  {volume} {78}},\ \bibinfo {pages} {123706} (\bibinfo {year}
  {2009})}\BibitemShut {NoStop}%
\bibitem [{\citenamefont {{Y. Xia, D. Qian, L. Wray, D. Hsieh, G. F. Chen, J.
  L. Luo, N. L. Wang and M. Z. Hasan}}(2009)}]{Y_XiaPRL2009}%
  \BibitemOpen
  \bibfield  {author} {\bibinfo {author} {\bibnamefont {{Y. Xia, D. Qian, L.
  Wray, D. Hsieh, G. F. Chen, J. L. Luo, N. L. Wang and M. Z. Hasan}}},\
  }\href@noop {} {\bibfield  {journal} {\bibinfo  {journal} {Phys. Rev. Lett},\
  }\textbf {\bibinfo {volume} {103}},\ \bibinfo {pages} {037002} (\bibinfo
  {year} {2009})}\BibitemShut {NoStop}%
\bibitem [{\citenamefont {{B. Mansart, V. Brouet, E. Papalazarou, M. Fuglsang
  Jensen, L. Petaccia, S. Gorovikov, A. N. Grum-Grzhimailo, F.
  Rullier-Albenque, A. Forget, D. Colson and M.
  Marsi}}(2011)}]{MansartPRB2011}%
  \BibitemOpen
  \bibfield  {author} {\bibinfo {author} {\bibnamefont {{B. Mansart, V. Brouet,
  E. Papalazarou, M. Fuglsang Jensen, L. Petaccia, S. Gorovikov, A. N.
  Grum-Grzhimailo, F. Rullier-Albenque, A. Forget, D. Colson and M. Marsi}}},\
  }\href@noop {} {\bibfield  {journal} {\bibinfo  {journal} {Phys. Rev. B},\
  }\textbf {\bibinfo {volume} {83}},\ \bibinfo {pages} {064516} (\bibinfo
  {year} {2011})}\BibitemShut {NoStop}%
\bibitem [{\citenamefont {{I. Nishi, M. Ishikado, S. Ideta, W. Malaeb, T.
  Yoshida, A. Fujimori, Y. Kotani, M. Kubota, K. Ono, M. Yi, D. H. Lu, R.
  Moore, Z.-X. Shen, A. Iyo, K. Kihou, H. Kito, H. Eisaki, S. Shamoto and R.
  Arita}}(2011)}]{Nishi_PRB2011}%
  \BibitemOpen
  \bibfield  {author} {\bibinfo {author} {\bibnamefont {{I. Nishi, M. Ishikado,
  S. Ideta, W. Malaeb, T. Yoshida, A. Fujimori, Y. Kotani, M. Kubota, K. Ono,
  M. Yi, D. H. Lu, R. Moore, Z.-X. Shen, A. Iyo, K. Kihou, H. Kito, H. Eisaki,
  S. Shamoto and R. Arita}}},\ }\href@noop {} {\bibfield  {journal} {\bibinfo
  {journal} {Phys. Rev. B},\ }\textbf {\bibinfo {volume} {84}},\ \bibinfo
  {pages} {014504} (\bibinfo {year} {2011})}\BibitemShut {NoStop}%
\bibitem [{\citenamefont {{Y. Zhang, F. Chen, C. He, B. Zhou, B. P. Xie, C.
  Fang, W. F. Tsai, X. H. Chen, H. Hayashi, J. Jiang, H. Iwasawa, K. Shimada,
  H. Namatame, M. Taniguchi, J. P. Hu, and D. L.
  Feng}}(2011)}]{Y_Zhang_PRB2011}%
  \BibitemOpen
  \bibfield  {author} {\bibinfo {author} {\bibnamefont {{Y. Zhang, F. Chen, C.
  He, B. Zhou, B. P. Xie, C. Fang, W. F. Tsai, X. H. Chen, H. Hayashi, J.
  Jiang, H. Iwasawa, K. Shimada, H. Namatame, M. Taniguchi, J. P. Hu, and D. L.
  Feng}}},\ }\href@noop {} {\bibfield  {journal} {\bibinfo  {journal} {Phys.
  Rev. B},\ }\textbf {\bibinfo {volume} {83}},\ \bibinfo {pages} {054510}
  (\bibinfo {year} {2011})}\BibitemShut {NoStop}%
\bibitem [{\citenamefont {{D. Venus}}(1993)}]{Venus_PRB1993}%
  \BibitemOpen
  \bibfield  {author} {\bibinfo {author} {\bibnamefont {{D. Venus}}},\
  }\href@noop {} {\bibfield  {journal} {\bibinfo  {journal} {Phys. Rev. B},\
  }\textbf {\bibinfo {volume} {48}},\ \bibinfo {pages} {6144} (\bibinfo {year}
  {1993})}\BibitemShut {NoStop}%
\bibitem [{\citenamefont {{A. Kaminski, S. Rosenkranz, H. M. Fretwell, J. C.
  Campuzano, Z. Li, H. Raffy, W. G. Cullen, H. You, C. G. Olsonk, C. M. Varma
  and H. H\"{o}chst}}(2002)}]{Kaminski_Nature2002}%
  \BibitemOpen
  \bibfield  {author} {\bibinfo {author} {\bibnamefont {{A. Kaminski, S.
  Rosenkranz, H. M. Fretwell, J. C. Campuzano, Z. Li, H. Raffy, W. G. Cullen,
  H. You, C. G. Olsonk, C. M. Varma and H. H\"{o}chst}}},\ }\href@noop {}
  {\bibfield  {journal} {\bibinfo  {journal} {Nature},\ }\textbf {\bibinfo
  {volume} {416}},\ \bibinfo {pages} {610} (\bibinfo {year}
  {2002})}\BibitemShut {NoStop}%
\bibitem [{\citenamefont {{K. Nakayama, T. Sato, K. Terashima, H. Matsui, T.
  Takahashi, M. Kubota, K. Ono, T. Nishizaki, Y. Takahashi and N.
  Kobayashi}}(2007)}]{Nakayama_PRB2007}%
  \BibitemOpen
  \bibfield  {author} {\bibinfo {author} {\bibnamefont {{K. Nakayama, T. Sato,
  K. Terashima, H. Matsui, T. Takahashi, M. Kubota, K. Ono, T. Nishizaki, Y.
  Takahashi and N. Kobayashi}}},\ }\href@noop {} {\bibfield  {journal}
  {\bibinfo  {journal} {Phys. Rev. B},\ }\textbf {\bibinfo {volume} {75}},\
  \bibinfo {pages} {014513} (\bibinfo {year} {2007})}\BibitemShut {NoStop}%
\bibitem [{\citenamefont {{V. B. Zabolotnyy, S. V. Borisenko, A. A. Kordyuk, J.
  Geck, D. S. Inosov, A. Koitzsch, J. Fink, M. Knupfer, B. B\"{u}chner, S.-L.
  Drechsler, H. Berger, A. Erb, M. Lambacher, L. Patthey, V. Hinkov and B.
  Keimer}}(2007)}]{Zabolotnyy_PRB2007_2}%
  \BibitemOpen
  \bibfield  {author} {\bibinfo {author} {\bibnamefont {{V. B. Zabolotnyy, S.
  V. Borisenko, A. A. Kordyuk, J. Geck, D. S. Inosov, A. Koitzsch, J. Fink, M.
  Knupfer, B. B\"{u}chner, S.-L. Drechsler, H. Berger, A. Erb, M. Lambacher, L.
  Patthey, V. Hinkov and B. Keimer}}},\ }\href@noop {} {\bibfield  {journal}
  {\bibinfo  {journal} {Phys. Rev. B},\ }\textbf {\bibinfo {volume} {76}},\
  \bibinfo {pages} {064519} (\bibinfo {year} {2007})}\BibitemShut {NoStop}%
\bibitem [{\citenamefont {{K. Nakayama, T. Sato, K. Terashima, T. Arakane, T.
  Takahashi, M. Kubota, K. Ono, T. Nishizaki, Y. Takahashi and N.
  Kobayashi}}(2009)}]{Nakayama_PRB2009}%
  \BibitemOpen
  \bibfield  {author} {\bibinfo {author} {\bibnamefont {{K. Nakayama, T. Sato,
  K. Terashima, T. Arakane, T. Takahashi, M. Kubota, K. Ono, T. Nishizaki, Y.
  Takahashi and N. Kobayashi}}},\ }\href@noop {} {\bibfield  {journal}
  {\bibinfo  {journal} {Phys. Rev. B},\ }\textbf {\bibinfo {volume} {79}},\
  \bibinfo {pages} {140503(R)} (\bibinfo {year} {2009})}\BibitemShut {NoStop}%
\bibitem [{\citenamefont {{V. B. Zabolotnyy, S. V. Borisenko, A. A. Kordyuk, D.
  S. Inosov, A. Koitzsch, J. Geck, J. Fink, M. Knupfer, B. B\"{u}chner, S.-L.
  Drechsler, V. Hinkov, B. Keimer and L.
  Patthey}}(2007)}]{Zabolotnyy_PRB2007_1}%
  \BibitemOpen
  \bibfield  {author} {\bibinfo {author} {\bibnamefont {{V. B. Zabolotnyy, S.
  V. Borisenko, A. A. Kordyuk, D. S. Inosov, A. Koitzsch, J. Geck, J. Fink, M.
  Knupfer, B. B\"{u}chner, S.-L. Drechsler, V. Hinkov, B. Keimer and L.
  Patthey}}},\ }\href@noop {} {\bibfield  {journal} {\bibinfo  {journal} {Phys.
  Rev. B},\ }\textbf {\bibinfo {volume} {76}},\ \bibinfo {pages} {024502}
  (\bibinfo {year} {2007})}\BibitemShut {NoStop}%
\bibitem [{\citenamefont {{H. Ding, P. Richard, K. Nakayama, K. Sugawara, T.
  Arakane, Y. Sekiba, A. Takayama, S. Souma, T. Sato, T. Takahashi, Z. Wang, X.
  Dai, Z. Fang, G. F. Chen, J. L. Luo and N. L. Wang}}(2008)}]{Ding_EPL2008}%
  \BibitemOpen
  \bibfield  {author} {\bibinfo {author} {\bibnamefont {{H. Ding, P. Richard,
  K. Nakayama, K. Sugawara, T. Arakane, Y. Sekiba, A. Takayama, S. Souma, T.
  Sato, T. Takahashi, Z. Wang, X. Dai, Z. Fang, G. F. Chen, J. L. Luo and N. L.
  Wang}}},\ }\href@noop {} {\bibfield  {journal} {\bibinfo  {journal}
  {Europhys. Lett.},\ }\textbf {\bibinfo {volume} {83}},\ \bibinfo {pages}
  {47001} (\bibinfo {year} {2008})}\BibitemShut {NoStop}%
\bibitem [{\citenamefont {{L. Zhao, H.-Y. Liu, W.-T. Zhang, J.-Q. Meng, X.-W.
  Jia, G.-D. Liu, X.-Li Dong, G.-F. Chen, J.-L. Luo, N.-L. Wang, W. Lu, G.-L.
  Wang, Y. Zhou, Y. Zhu, X.-Y. Wang, Z.-Y. Xu, C.-T. Chen and X.-J.
  Zhou}}(2008)}]{L_ZhaoCPL2008}%
  \BibitemOpen
  \bibfield  {author} {\bibinfo {author} {\bibnamefont {{L. Zhao, H.-Y. Liu,
  W.-T. Zhang, J.-Q. Meng, X.-W. Jia, G.-D. Liu, X.-Li Dong, G.-F. Chen, J.-L.
  Luo, N.-L. Wang, W. Lu, G.-L. Wang, Y. Zhou, Y. Zhu, X.-Y. Wang, Z.-Y. Xu,
  C.-T. Chen and X.-J. Zhou}}},\ }\href@noop {} {\bibfield  {journal} {\bibinfo
   {journal} {Chin. Phys. Lett.},\ }\textbf {\bibinfo {volume} {25}},\ \bibinfo
  {pages} {4402} (\bibinfo {year} {2008})}\BibitemShut {NoStop}%
\bibitem [{\citenamefont {{Y-M. Xu, Y-B. Huang, X-Y. Cui, E. Razzoli, M.
  Radovic, M. Shi, G-F. Chen, P. Zheng, N-L.Wang, C-L. Zhang, P-C. Dai, J-P.
  Hu, Z. Wang and H. Ding}}(2011)}]{YM_Xu_NPhys2011}%
  \BibitemOpen
  \bibfield  {author} {\bibinfo {author} {\bibnamefont {{Y-M. Xu, Y-B. Huang,
  X-Y. Cui, E. Razzoli, M. Radovic, M. Shi, G-F. Chen, P. Zheng, N-L.Wang, C-L.
  Zhang, P-C. Dai, J-P. Hu, Z. Wang and H. Ding}}},\ }\href@noop {} {\bibfield
  {journal} {\bibinfo  {journal} {Nature Phys.},\ }\textbf {\bibinfo {volume}
  {7}},\ \bibinfo {pages} {198} (\bibinfo {year} {2011})}\BibitemShut {NoStop}%
\bibitem [{\citenamefont {{S.Graser, T.A.Maier, P.J. Hirschfeld and D.J.
  Scalapino}}(2009)}]{Graser_NJP2009}%
  \BibitemOpen
  \bibfield  {author} {\bibinfo {author} {\bibnamefont {{S.Graser, T.A.Maier,
  P.J. Hirschfeld and D.J. Scalapino}}},\ }\href@noop {} {\bibfield  {journal}
  {\bibinfo  {journal} {New J. Phys.},\ }\textbf {\bibinfo {volume} {11}},\
  \bibinfo {pages} {025016} (\bibinfo {year} {2009})}\BibitemShut {NoStop}%
\bibitem [{\citenamefont {{S. Graser, A. F. Kemper, T. A. Maier, H.-P. Cheng,
  P. J. Hirschfeld and D. J. Scalapino}}(2010)}]{Graser_PRB2010}%
  \BibitemOpen
  \bibfield  {author} {\bibinfo {author} {\bibnamefont {{S. Graser, A. F.
  Kemper, T. A. Maier, H.-P. Cheng, P. J. Hirschfeld and D. J. Scalapino}}},\
  }\href@noop {} {\bibfield  {journal} {\bibinfo  {journal} {Phys. Rev. B},\
  }\textbf {\bibinfo {volume} {81}},\ \bibinfo {pages} {214503} (\bibinfo
  {year} {2010})}\BibitemShut {NoStop}%
\bibitem [{\citenamefont {{Chia-Hui Lin, Tom Berlijn, Limin Wang, Chi-Cheng
  Lee, Wei-Guo Yin and Wei Ku}}(2011)}]{CH_Lin_PRL2011}%
  \BibitemOpen
  \bibfield  {author} {\bibinfo {author} {\bibnamefont {{Chia-Hui Lin, Tom
  Berlijn, Limin Wang, Chi-Cheng Lee, Wei-Guo Yin and Wei Ku}}},\ }\href@noop
  {} {\bibfield  {journal} {\bibinfo  {journal} {Phys. Rev. Lett.},\ }\textbf
  {\bibinfo {volume} {107}},\ \bibinfo {pages} {257001} (\bibinfo {year}
  {2011})}\BibitemShut {NoStop}%
\bibitem [{\citenamefont {{P. A. Lee, X.-G. Wen}}(2008)}]{XGWen_3orbital}%
  \BibitemOpen
  \bibfield  {author} {\bibinfo {author} {\bibnamefont {{P. A. Lee, X.-G.
  Wen}}},\ }\href@noop {} {\bibfield  {journal} {\bibinfo  {journal} {Phys.
  Rev. B},\ }\textbf {\bibinfo {volume} {78}},\ \bibinfo {pages} {144517}
  (\bibinfo {year} {2008})}\BibitemShut {NoStop}%
\bibitem [{\citenamefont {{P. Zhang, P. Richard, T. Qian, Y.-M. Xu, X. Dai and
  H. Ding}}(2011)}]{Zhang_RSI2011}%
  \BibitemOpen
  \bibfield  {author} {\bibinfo {author} {\bibnamefont {{P. Zhang, P. Richard,
  T. Qian, Y.-M. Xu, X. Dai and H. Ding}}},\ }\href@noop {} {\bibfield
  {journal} {\bibinfo  {journal} {Rev. Sci. Instrum.},\ }\textbf {\bibinfo
  {volume} {82}},\ \bibinfo {pages} {043712} (\bibinfo {year}
  {2011})}\BibitemShut {NoStop}%
\bibitem [{\citenamefont {{Hai-Jun Zhang, Gang Xu, Xi Dai, Zhong
  Fang}}(2009)}]{Zhang_CPL2009}%
  \BibitemOpen
  \bibfield  {author} {\bibinfo {author} {\bibnamefont {{Hai-Jun Zhang, Gang
  Xu, Xi Dai, Zhong Fang}}},\ }\href@noop {} {\bibfield  {journal} {\bibinfo
  {journal} {Chin. Phys. Lett.},\ }\textbf {\bibinfo {volume} {26}},\ \bibinfo
  {pages} {017401} (\bibinfo {year} {2009})}\BibitemShut {NoStop}%
\bibitem [{\citenamefont {{Ver—nica Vildosola, Leonid Pourovskii, Ryotaro
  Arita, Silke Biermann, and Antoine Georges}}(2008)}]{Vildosola_2008}%
  \BibitemOpen
  \bibfield  {author} {\bibinfo {author} {\bibnamefont {{Ver—nica Vildosola,
  Leonid Pourovskii, Ryotaro Arita, Silke Biermann, and Antoine Georges}}},\
  }\href@noop {} {\bibfield  {journal} {\bibinfo  {journal} {Phys. Rev. B},\
  }\textbf {\bibinfo {volume} {78}},\ \bibinfo {pages} {064518} (\bibinfo
  {year} {2008})}\BibitemShut {NoStop}%
\bibitem [{\citenamefont {{A. Damascelli}}(2004)}]{DamascelliPScrypta2004}%
  \BibitemOpen
  \bibfield  {author} {\bibinfo {author} {\bibnamefont {{A. Damascelli}}},\
  }\href@noop {} {\bibfield  {journal} {\bibinfo  {journal} {Phys. Scrypta},\
  }\textbf {\bibinfo {volume} {T109}},\ \bibinfo {pages} {61} (\bibinfo {year}
  {2004})}\BibitemShut {NoStop}%
\bibitem [{\citenamefont {{J. J. Yeh and I. Lindau}}(1985)}]{Yeh1985}%
  \BibitemOpen
  \bibfield  {author} {\bibinfo {author} {\bibnamefont {{J. J. Yeh and I.
  Lindau}}},\ }\href@noop {} {\bibfield  {journal} {\bibinfo  {journal} {At.
  Data Nucl. Data Tables},\ }\textbf {\bibinfo {volume} {32}},\ \bibinfo
  {pages} {1} (\bibinfo {year} {1985})}\BibitemShut {NoStop}%
\bibitem [{\citenamefont {{H. Miao, P. Richard, Y. Tanaka, K. Nakayama, T.
  Qian, K. Umezawa, T. Sato, Y.-M. Xu, Y.-B. Shi, N. Xu, X.-P. Wang, P. Zhang,
  H.-B. Yang, Z.-J. Xu, J. S. Wen, G.-D. Gu, X. Dai, J.-P. Hu, T. Takahashi and
  H. Ding}}(2012)}]{H_Miao}%
  \BibitemOpen
  \bibfield  {author} {\bibinfo {author} {\bibnamefont {{H. Miao, P. Richard,
  Y. Tanaka, K. Nakayama, T. Qian, K. Umezawa, T. Sato, Y.-M. Xu, Y.-B. Shi, N.
  Xu, X.-P. Wang, P. Zhang, H.-B. Yang, Z.-J. Xu, J. S. Wen, G.-D. Gu, X. Dai,
  J.-P. Hu, T. Takahashi and H. Ding}}},\ }\href@noop {} {\bibfield  {journal}
  {\bibinfo  {journal} {Phys. Rev. B},\ }\textbf {\bibinfo {volume} {85}},\
  \bibinfo {pages} {094506} (\bibinfo {year} {2012})}\BibitemShut {NoStop}%
\bibitem [{\citenamefont {{J. Kang and Z. Tesanovic}}(2011)}]{J_KangPRB2011}%
  \BibitemOpen
  \bibfield  {author} {\bibinfo {author} {\bibnamefont {{J. Kang and Z.
  Tesanovic}}},\ }\href@noop {} {\bibfield  {journal} {\bibinfo  {journal}
  {Phys. Rev. B},\ }\textbf {\bibinfo {volume} {83}},\ \bibinfo {pages}
  {020505(R)} (\bibinfo {year} {2011})}\BibitemShut {NoStop}%
\bibitem [{\citenamefont {{K. Nakayama, T. Sato, P. Richard, Y.-M. Xu, Y.
  Sekiba, S. Souma, G. F. Chen, J. L. Luo, N. L. Wang, H. Ding and T.
  Takahashi}}(2009)}]{Nakayama_EPL2009}%
  \BibitemOpen
  \bibfield  {author} {\bibinfo {author} {\bibnamefont {{K. Nakayama, T. Sato,
  P. Richard, Y.-M. Xu, Y. Sekiba, S. Souma, G. F. Chen, J. L. Luo, N. L. Wang,
  H. Ding and T. Takahashi}}},\ }\href@noop {} {\bibfield  {journal} {\bibinfo
  {journal} {Europhys. Lett.},\ }\textbf {\bibinfo {volume} {85}},\ \bibinfo
  {pages} {67002} (\bibinfo {year} {2009})}\BibitemShut {NoStop}%
\bibitem [{\citenamefont {{K. Nakayama, T. Sato, P. Richard, Y.-M. Xu, T.
  Kawahara, K. Umezawa, T. Qian, M. Neupane, G. F. Chen, H. Ding and T.
  Takahashi}}(2011)}]{Nakayama_PRB2011}%
  \BibitemOpen
  \bibfield  {author} {\bibinfo {author} {\bibnamefont {{K. Nakayama, T. Sato,
  P. Richard, Y.-M. Xu, T. Kawahara, K. Umezawa, T. Qian, M. Neupane, G. F.
  Chen, H. Ding and T. Takahashi}}},\ }\href@noop {} {\bibfield  {journal}
  {\bibinfo  {journal} {Phys. Rev. B},\ }\textbf {\bibinfo {volume} {83}},\
  \bibinfo {pages} {020501(R)} (\bibinfo {year} {2011})}\BibitemShut {NoStop}%
\bibitem [{\citenamefont {{Y.-M. Xu, P. Richard, K. Nakayama, T. Kawahara, Y.
  Sekiba, T. Qian, M. Neupane, S. Souma, T. Sato, T. Takahashi, H.-Q. Luo,
  H.-H. Wen, G.-F. Chen, N.-L. Wang, Z. Wang, Z. Fang, X. Dai and H.
  Ding}}(2011)}]{YM_Xu_Ncommun2011}%
  \BibitemOpen
  \bibfield  {author} {\bibinfo {author} {\bibnamefont {{Y.-M. Xu, P. Richard,
  K. Nakayama, T. Kawahara, Y. Sekiba, T. Qian, M. Neupane, S. Souma, T. Sato,
  T. Takahashi, H.-Q. Luo, H.-H. Wen, G.-F. Chen, N.-L. Wang, Z. Wang, Z. Fang,
  X. Dai and H. Ding}}},\ }\href@noop {} {\bibfield  {journal} {\bibinfo
  {journal} {Nat. Commun.},\ }\textbf {\bibinfo {volume} {2}},\ \bibinfo
  {pages} {392} (\bibinfo {year} {2011})}\BibitemShut {NoStop}%
\bibitem [{\citenamefont {{Z.-H. Liu, P. Richard, K. Nakayama, G.-F. Chen, S.
  Dong, J.-B. He, D.-M. Wang, T.-L. Xia, K. Umezawa, T. Kawahara, S. Souma, T.
  Sato, T. Takahashi, T. Qian, Yaobo Huang, Nan Xu, Yingbo Shi, H. Ding and
  S.-C. Wang}}(2011)}]{ZH_LiuPRB2011}%
  \BibitemOpen
  \bibfield  {author} {\bibinfo {author} {\bibnamefont {{Z.-H. Liu, P. Richard,
  K. Nakayama, G.-F. Chen, S. Dong, J.-B. He, D.-M. Wang, T.-L. Xia, K.
  Umezawa, T. Kawahara, S. Souma, T. Sato, T. Takahashi, T. Qian, Yaobo Huang,
  Nan Xu, Yingbo Shi, H. Ding and S.-C. Wang}}},\ }\href@noop {} {\bibfield
  {journal} {\bibinfo  {journal} {Phys. Rev. B},\ }\textbf {\bibinfo {volume}
  {84}},\ \bibinfo {pages} {064519} (\bibinfo {year} {2011})}\BibitemShut
  {NoStop}%
\bibitem [{\citenamefont {{J.-P. Hu and H. Ding}}(2011)}]{JP_Hu2011}%
  \BibitemOpen
  \bibfield  {author} {\bibinfo {author} {\bibnamefont {{J.-P. Hu and H.
  Ding}}},\ }\href@noop {} {\bibfield  {journal} {\bibinfo  {journal}
  {arXiv:1107.1334v1}} (\bibinfo {year} {2011})}\BibitemShut {NoStop}%
\end{thebibliography}%

\end{document}